\documentclass[11pt]{article}
\pdfoutput=1  
\usepackage{graphicx}
\graphicspath{{./figures/}}
\usepackage{appendix}
\usepackage{latexsym,amsmath,amsfonts,amssymb,booktabs}
\usepackage[font=small]{caption}
\usepackage{slashed,upgreek,amscd,cancel,tensor,color}
\usepackage{adjustbox}
\usepackage[numbers,compress,square]{natbib}
\usepackage{epsfig,latexsym}
\usepackage[pdfencoding=auto]{hyperref}
\usepackage{url}
\numberwithin{equation}{section}
\usepackage{doi}
\usepackage{subcaption}
\definecolor{MyBlue}{rgb}{0.15,0.15,0.70}

\hypersetup{
colorlinks=true,
citecolor=MyBlue,
linkcolor=MyBlue,
urlcolor=MyBlue
}

\input epsf
\usepackage{amssymb}
\usepackage{amsmath}
\usepackage{amsfonts}
\usepackage{upgreek}
\usepackage{latexsym}
\usepackage{stfloats}
\usepackage{afterpage}

\newcommand{\be}{\begin{equation}}
\newcommand{\ee}{\end{equation}}
\newcommand{\beq}{\begin{equation}}
\newcommand{\eeq}{\end{equation}}
\newcommand{\bea}{\begin{eqnarray}}
\newcommand{\eea}{\end{eqnarray}}

\newcommand{\R}{R}

\usepackage[stable]{footmisc}

\def\d{\delta}

\def\MM{M_{*}}

\def\dkmu2{\delta K_{\mu \nu}\delta K^{\mu \nu}}
\def\pmu2{  \phi_{\mu \nu}\phi^{\mu \nu}}

\newcommand\bone{{\alpha}_{\rm V1}}
\newcommand\btwo{{\alpha}_{\rm V2}}
\newcommand\bthree{{\alpha}_{\rm V3}}

\newcommand\dotbone{\dot \alpha_{\rm V1} }
\newcommand\dotbtwo{\dot \alpha_{\rm V2} }
\newcommand\dotbthree{\dot \alpha_{\rm V3} }

\newcommand\alphaq{{\alpha}_{\rm V1}}
\newcommand\alphac{{\alpha}_{\rm V2}}

\newcommand\alphaset{{\alpha}_{\rm V3}}

\newcommand\T{{\cal T}}
\newcommand\SSS{{\cal S}}

\newcommand{\deltac}{\delta}
\newcommand{\AAA}{A_1}
\newcommand{\BBB}{A_2}
\newcommand{\CCC}{A_3}
\newcommand{\AAAA}{A}

\newcommand{\gmn}{g_{\mu\nu}}

\renewcommand\[{\left[}

\newcommand\ees{\end{eqnarray}}
\newcommand\bees{\begin{eqnarray}}

\newcommand\alphaB{\alpha_{\text{B}}}
\newcommand\alphaM{\alpha_{\text{M}}}

\newcommand\alphaT{\alpha_{\text{T}}}

\newcommand{\unitsk}{ h \, \text{Mpc}^{-1}}
\newcommand{\kvec}{\vec{k}}
\newcommand{\qvec}{\vec{q}}
\newcommand{\cH}{\mathcal{H}}
\newcommand{\eqn}[1]{eq.~(\ref{#1})}

 \newcommand{\vk}{\vec{k}}
 
 \newcommand{\ta}{\tilde{a}}
 \newcommand{\momspmeas}[1]{\frac{d^3 #1}{(2 \pi)^3}}
 \newcommand{\half}{\frac{1}{2}}
 \newcommand{\muphi}{\mu_{\Phi}}
 \newcommand{\mupsi}{\mu_{\Psi}}
 \newcommand{\muchi}{\mu_{\chi}}

\newcommand{\omegam}{\Omega_{\rm m}}
\newcommand{\omegamz}{\Omega_{\rm m, 0}}

\newcommand{\xvec}{\vec{x}}
\newcommand{\knl}{k_{\rm NL}}
\newcommand{\deltam}{\delta}
\newcommand{\rhom}{\rho_{\rm m}}
\newcommand{\dotrhom}{{\dot \rho}_{\rm m}}

\newcommand{\deltan}{\delta^{(n)}}
\newcommand{\thetan}{\Theta^{(n)}}

\newcommand{\mpl}{M_{\rm Pl}}

\begin{document}
\vspace{0.5cm}

\begin{center}
\Large{\textbf{ Dark Energy and Modified Gravity in  \\
the Effective Field Theory of Large-Scale Structure}} \\[1cm]

\large{Giulia Cusin$^a$, Matthew Lewandowski$^b$ and Filippo Vernizzi$^b$}
\\[0.5cm]

\small{
\textit{$^a$ D\'epartement de Physique Th\'eorique and Center for Astroparticle Physics, \\
Universit\'e de Gen\`eve, 24 quai Ansermet, CH--1211 Gen\`eve 4, Switzerland}}

\vspace{.2cm}

\small{
\textit{$^b$ Institut de physique th\' eorique, Universit\'e  Paris Saclay \\ [0.05cm]
CEA, CNRS, 91191 Gif-sur-Yvette, France  }}

\vspace{.2cm}

\vspace{0.5cm}
\today

\end{center}

\vspace{2cm}

\begin{abstract}

{We develop an approach to compute observables beyond the linear regime of dark matter perturbations for general dark energy and modified gravity models. We do so by combining the  Effective Field Theory of Dark Energy and  Effective Field Theory of Large-Scale Structure approaches.
In particular, we parametrize the linear and nonlinear effects of dark energy on dark matter clustering in terms of the Lagrangian terms introduced in a companion paper \cite{CLV1}, focusing on Horndeski theories and assuming the quasi-static approximation.
The Euler equation for dark matter is  sourced, via the Newtonian potential,  by new nonlinear vertices due to modified gravity 
and, as in the pure dark matter  case, by the effects of short-scale physics in the form of the divergence of an effective stress tensor. 
The effective fluid introduces a counterterm in the  solution to the matter continuity and Euler equations,
which allows  a controlled expansion of clustering statistics on mildly nonlinear scales.  
We use this setup to compute the one-loop dark-matter power spectrum.}

\end{abstract}

\newpage

\tableofcontents

\vspace{.5cm}
\newpage

\section{Introduction}

In the near future, large-scale structure (LSS) surveys have the chance to remarkably increase our understanding of the recent universe by measuring its expansion history and the properties of the clustering of massive objects.  While the properties of gravity are highly constrained in the early universe and on solar system scales, the new wealth of data will allow us to test gravity on scales where, so far, much less is known.  This gives us the opportunity to probe in detail various cosmological scenarios, including dark energy and modified gravity theories, which could leave observable signatures in upcoming LSS surveys (see e.g.~\cite{Amendola:2012ys,Amendola:2016saw} and references therein).  Given that we will have such precise data, we are pressed to understand how to use it in the best way.  

Because most of the aforementioned data will be concentrated on short scales where gravitational nonlinearities become large, one must understand the mildly nonlinear regime of structure formation.  This challenge has already been recognized in the attempt to constrain primordial non-Gaussianity (see e.g.~\cite{Alvarez:2014vva} and references therein) with LSS measurements, so that much work has already been done to understand the mildly nonlinear regime (see e.g.~\cite{Crocce:2005xy,Bernardeau:2008fa,Bernardeau:2011vy,McDonald:2006mx,Matarrese:2007wc,Taruya:2007xy,Matsubara:2007wj,Pietroni:2008jx,Baumann:2010tm,Carrasco:2012cv,Bartelmann:2014qca,Blas:2015qsi} for a non-exhaustive list).  In the case of primordial non-Gaussianity, strong constraints have already been made by the cosmic microwave background (CMB), and so one must very precisely understand LSS observables in order to use them to make improved constraints.  The situation is more promising, however, for dark energy and modified gravity.  Because they are low-redshift phenomena, they are currently much less constrained by the cosmic microwave background radiation, so it is expected that LSS measurements will play a more important role in their understanding.

{This  is the second of a series of two papers.  In the first one  \cite{CLV1}, we constructed the nonlinear action for dark energy and modified gravity theories characterized by a single scalar degree of freedom,
using the language of the Effective Field Theory of Dark Energy (EFTofDE) \cite{Creminelli:2008wc,Gubitosi:2012hu,Bloomfield:2012ff,Gleyzes:2013ooa,Bloomfield:2013efa}. An important advantage of this approach  is that it ensures that  predictions are consistent with well-established physical principles such as locality, causality, stability and unitarity, because these conditions can be imposed at the level of the Lagrangian.  

In this article, we use the EFTofDE to parametrize  the effect of  dark energy and modified gravity  on linear and nonlinear perturbations in the quasi-static regime. Moreover, we model the gravitational clustering of dark matter in the mildly nonlinear regime using the Effective Field Theory of Large-Scale Structure (EFTofLSS) approach \cite{Baumann:2010tm,Carrasco:2012cv}. For the inclusion of a clustering quintessence type dark energy \cite{Creminelli:2009mu,Sefusatti:2011cm} in the EFTofLSS, see \cite{Lewandowski:2016yce}.}

In order to correctly describe gravitational clustering at the highest wavenumbers possible, the EFTofLSS was developed to correctly treat, in perturbation theory, the effects of short-scale modes on the long-wavelength observables measured in LSS.  The central idea of this approach is to include appropriate counterterms (which have free coefficients not predictable within the EFT) in the perturbative expansion, which systematically correct mistakes introduced in loops from uncontrolled short-distance physics.  One is then left with a controlled expansion in $k / \knl$, where $\knl$ is the strong coupling scale of the EFT (i.e. the EFT can not describe scales above $\knl$ due to unknown UV effects).  

This has at least three important benefits.  First, the maximum wavenumber describable with the theory 
has been increased over former analytic treatments \cite{Foreman:2015lca}.  Second, for $ k  \lesssim \knl$, observables can be computed to higher and higher precision by including more and more loops (up to non-perturbative effects).  Finally, one is able to estimate the theoretical error in any computation by estimating the size of the next loop contribution which has not been computed.  So far, this research effort has shown that clustering can be accurately described for dark matter bispectrum \cite{Angulo:2014tfa,Baldauf:2014qfa}, one-loop trispectrum \cite{Bertolini:2016bmt}, galaxies \cite{Senatore:2014eva, Angulo:2015eqa,Assassi:2015fma}, including baryons \cite{Lewandowski:2014rca} and massive neutrinos \cite{Senatore:2017hyk}, for the BAO peak \cite{Senatore:2014via, Baldauf:2015xfa},  in redshift space \cite{Perko:2016puo, Lewandowski:2015ziq, delaBella:2017qjy}, and including primordial non-Gaussianity \cite{Assassi:2015jqa,Assassi:2015fma}.  Importantly, now that the theory is on a firm footing, research has been able to move on to the practical question of efficient numerical implementation of the above ideas \cite{Cataneo:2016suz, Simonovic:2017mhp}.  In this article, we would like to start using the machinery of the EFTofLSS to develop a robust way to test nonlinearities in dark energy and modified gravity theories.

{First, in Sec.~\ref{reviewsec}, we review the development of the nonlinear EFTofDE presented in  \cite{CLV1}. 
In particular, we present the linear and nonlinear operators describing scalar-tensor theories in the Horndeski class, assuming the quasi-static regime. We then show how the modifications of gravity induced by the scalar field enter the LSS equations as a modified Poisson equation, relating the second derivative of the gravitational potential to a nonlinear combination of the matter overdensity.

Then, we move on to derive the equations for the dark-matter overdensity, described by the continuity and Euler equations.
The effect of unknown short-scale physics enters as a source term in  the Euler equation.
As reviewed in Sec.~\ref{Effective fluid}, in the standard dark matter case this source term has the form of a total spatial derivative of an effective stress-energy tensor. Using the EFTofDE approach we show that, in the quasi-static non-relativistic limit, this is the case also in the presence of dark energy and modified gravity.
Therefore, at one-loop in perturbations this term gives rise to a counterterm that enters the dark-matter equations in the same way as in general relativity.}

In Sec.~\ref{Perturbation theory} we use the perturbation equations in the presence of dark energy and modified gravity to compute the power spectrum at  one-loop order.  
In our perturbative approach, we use the exact Green's functions of the linear equations (which are scale independent) to solve for the time dependence at higher orders.  
This is to be contrasted with the case in $\Lambda$CDM where one can normally use the Einstein de Sitter approximation, which is accurate to better than one percent \cite{Takahashi:2008yk} for the time dependence of higher loop terms. Moreover, we include the effect of the counterterms due to the short-scale modes, which in our calculation is a free parameter. The details of the calculation and several definitions are reported in App. \ref{Apponeloop}.
{For other one-loop calculations in perturbation theory including the effect of modified gravity see e.g.~\cite{Koyama:2009me,Takushima:2013foa,Takushima:2015iha,Song:2015oza,Bose:2016qun}.}

In order to compute the perturbative expansion, one must perform loop integrals over intermediate momenta.  
As discussed for instance in \cite{Jain:1995kx, Scoccimarro:1995if,Bernardeau:2011vy, Bernardeau:2012aq,Peloso:2013zw,Kehagias:2013yd}, individual terms in the loop expansion that are summed to compute the equal-time power spectrum up to one-loop may contain spurious infrared (IR) divergences. These ultimately must cancel in the final expression, because of the equivalence principle~\cite{Carrasco:2013sva, Creminelli:2013mca}.  
A similar cancellation takes place in the ultraviolet (UV) because of matter and momentum conservation. 
However, because in general  the different contributions have different time dependences, a small numerical error in the calculation of their  coefficients may lead to an incomplete cancellation~\cite{Lewandowski:2017kes}.  
This motivates us to use the IR\&UV-safe versions of the momentum integrals, where  divergences are subtracted out at the level of the integrand \cite{Carrasco:2013sva,Blas:2013bpa, Lewandowski:2017kes}.  We explicitly show that the spurious IR terms cancel at one loop, also including modifications of gravity, as expected from the fact that the equivalence principle is satisfied in our case. We report the details of this technical but important issue in App. \ref{irsafesec}. {A consequence of this finding is that the so-called consistency relations for LSS \cite{Peloso:2013zw,Kehagias:2013yd,Creminelli:2013mca,Creminelli:2013poa} are satisifed also in our case.}

\begin{figure}[htb!]
\centering
\begin{subfigure}{4.1in}
\includegraphics[width=10cm]{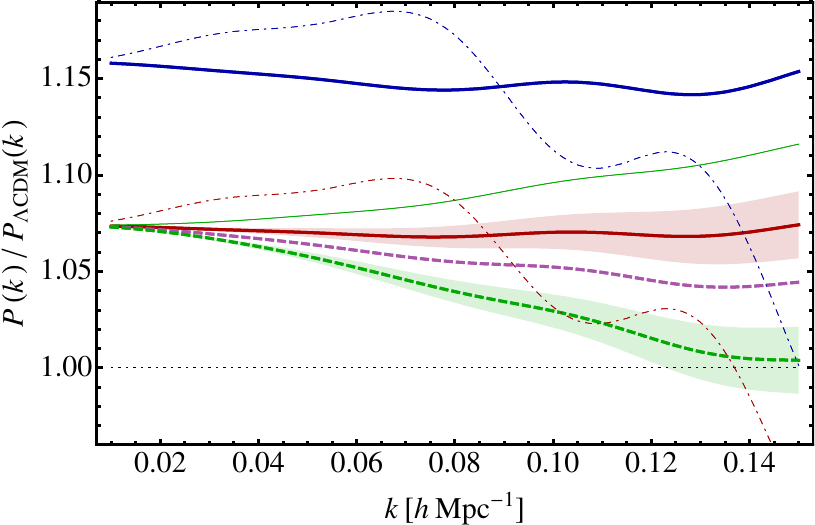} 
\end{subfigure}
\begin{subfigure}{2in}
  \includegraphics[width=5cm]{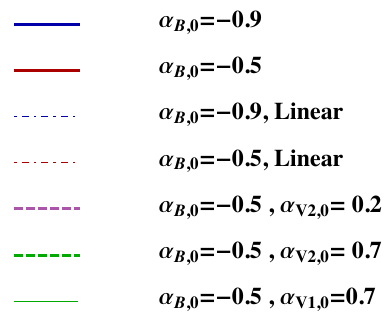}
\end{subfigure}
\caption{ \footnotesize  Effect of some of the modified gravity couplings on the one-loop matter power spectrum.  The ratio between the predicted  up-to-one-loop power spectrum  with dark energy and that for $\Lambda$CDM is shown for different current values of three modified gravity parameters, $\alphaB $, $\alphaq$, and $\alphac $, including LSS counterterms.  Specifically, $\alphaB $  enters at quadratic and higher order  in the action while $\alphaq$ and $\alphac $ enter at cubic and higher order, so they do not modify the linear spectrum. All of the modified  gravity parameters which are not mentioned in the legend are set to zero.  Moreover, the curves labeled as ``Linear'' (thin dashed-dot lines) are the linear predictions for the corresponding values of $\alpha_{\rm B,0}$.
The bands around the dashed green and red curves are obtained by varying the amplitude of the LSS counterterm over a reasonable range (see more details in Sec.~\ref{resultssec}).  
Since  a non-vanishing $\alphaB$ changes also the linear power spectrum with dark energy, large modifications on mildly nonlinear scales due to this parameter also imply large changes in the linear spectrum. On the other hand, $\alphaq$  and $\alphac$  have a direct  effect on mildly-nonlinear scales without affecting  the linear predictions.   } \label{alphabplot}
\end{figure}

To anticipate some of our results, in Fig.~\ref{alphabplot} we show a small sample of the effects of modified gravity on the one-loop power spectrum. In particular, we show  the one-loop power spectrum computed by turning on three different couplings, one that starts at quadratic order in the action, $\alphaB$, and two that start at cubic order in the action, $\alphaq$ and $\alphac$ (see Sec. \ref{reviewsec}), divided by the one-loop power spectrum in $\Lambda$CDM.

As expected, the main effect of $\alphaB$ is to change the linear power spectrum.  This can be seen in the low $k$ behavior of the solid red and solid blue curves in Fig.~\ref{alphabplot}.  The nonlinear effects from $\alphaB$, which appear as modifications at higher $k$, are only substantial if they are accompanied by a large change in the linear power spectrum.  On the other hand, by turning on the nonlinear couplings $\alphaq$ or $\alphac$, we directly modify the power spectrum on nonlinear scales, as can be seen, for example, in the thick dashed green curve in Fig.~\ref{alphabplot}.  The green and red shaded regions are given by varying the value of the counterterm for the respective curves over a reasonable range.  More details on this plot can be found in Sec.~\ref{resultssec}.


%
%
%

\section{Nonlinear effective theory of dark energy} \label{reviewsec}
 
In this section, we briefly review the results from \cite{CLV1} that are relevant to compute the one-loop dark-matter power spectrum.  In the EFTofDE, the dark-energy field is the Goldstone mode of broken time diffeomorphisms {\cite{Creminelli:2006xe, Cheung:2007st}}.  Ultimately, this means that the action in unitary gauge can be written in terms of operators which are fully diffeomorphism invariant, like the 4-dimensional Ricci scalar $ {}^{(4)}R$, and operators that are invariant under the remaining time-dependent spatial diffeomorphisms.  These are tensors with upper zero indices, like $g^{00}$, the extrinsic curvature $K_{\mu \nu}$, and the 3-dimensional curvature $R_{\mu \nu}$ of the spatial metric $h_{\mu \nu}$, all of which can have arbitrary time-dependent coefficients.  

{It is convenient to define the following combinations of  operators,} 
\be\label{K2}
\begin{split}
\delta {\cal K}_2 & \equiv  \delta K^2 - \d K^\nu{}_\mu \d K^\mu{}_\nu  \;, \qquad
\delta {\cal G}_2  \equiv \delta K^\nu{}_\mu \R^\mu{}_\nu - \frac12 \d K \R \;, \\
\end{split}
\ee
\be\label{K3}
\delta {\cal K}_3  \equiv \delta K^3 -3 \, \d K \d K^\nu{}_\mu \, \d K^\mu{}_\nu+ 2 \d K^\nu{}_\mu \,  \d K^\mu{}_\rho \, \d K^\rho{}_\nu \; ,
\ee
where $\delta K_{\mu \nu} = K_{\mu \nu} - H h_{\mu \nu}$, $K = h^{\mu \nu} K_{\mu \nu}$, and $R = h^{\mu \nu} R_{\mu \nu}$.  
Focusing on Horndeski theories \cite{Horndeski:1974wa,Deffayet:2011gz} and restricting to the  quasi-static, non-relativistic limit,  the full action in unitary gauge for the gravitational sector is given by \cite{CLV1}
\be
\begin{split}
\label{total_actionEFT}
S_{\rm g}  = & \int  d^4 x \sqrt{-g}  \bigg\{ \frac{\MM^2 f (t)}{2}  {}^{(4)}\!R  - \frac{m_3^3(t)}{2} \, \delta K \delta g^{00}  
- m_4^2 (t) \left( \delta {\cal K}_2- \frac{ 1}{2} \,\delta g^{00} \R \right)    \\
 &    \hspace{1in} - \frac{m_5^2(t)}{2}  \delta g^{00} \delta {\cal K}_2-   \frac{m_6(t)}{3} (\delta {\cal K}_3   + 3  \delta g^{00} \delta {\cal G}_2)    -   \frac{m_7(t)}{3} \delta g^{00} \delta {\cal K}_3 \bigg\} \; , 
\end{split}
\ee
where $\delta g^{00} = 1 + g^{00}$.
The effective Planck mass (the coefficient in front of the time derivative part of the graviton kinetic term) is given by 
\be
M^2   \equiv \MM^2  f + 2 m_4^2 \;.
\ee
We can then define the dimensionless versions of the coefficients in \eqn{total_actionEFT} as \cite{Gleyzes:2014rba,CLV1}
\be
\begin{split}
\label{EFTaction_masses}
 \alphaB & \equiv \frac{\MM^2 \dot f - m_3^3 }{2 M^2 H} \;,  \qquad
 \alphaM  \equiv  \frac{ \MM^2 \dot f + 2 (m_4^2)^{\hbox{$\cdot$}}}{M^2 H }\;, \qquad  \alphaT  \equiv - \frac{2 m_4^2}{M^2 } \;,  \\
  \bone  & \equiv \frac{2 m_5^2+2 H m_6}{M^2} \;, \qquad  \btwo \equiv  \frac{2 H m_6}{M^2} \;, \qquad  \bthree \equiv  \frac{4 H  m_7+2 H m_6 }{M^2}  \;.
\end{split}
\ee
The dark energy field $\pi$, or its dimensionless version $ \chi \equiv H \pi$, can then be introduced into the action \eqref{total_actionEFT} by performing a time change of coordinates.  

The total action for our system, then is $S = S_{\rm g} + S_{\rm m}$, with 
\be \label{matteraction}
S_{\rm m} = - \frac12  \int d^4 x \sqrt{-g}  T^{(\rm m)}_{\mu \nu} \, \delta g^{\mu \nu} \;,
\ee 
where $T^{(\rm m)}_{\mu \nu}$ is the stress-energy tensor of cold dark matter, which we assume is minimally coupled to the metric $\gmn$.  In the long wavelength limit, dark matter is a non-relativistic perfect fluid with zero speed of sound, for which the stress-energy tensor is
\be \label{matterstresstensor}
T^{(\rm m)}{}^{ \mu}{}_\nu = \rho_{\rm m} u^\mu u_\nu \;, 
\ee
where $ \rho_{\rm m}$ is the energy density in the rest frame of the fluid and $u^\mu$ is the 4-velocity.  In the non-relativistic limit, we have  
$u^\mu =  (1/\sqrt{-g_{00}}, a^{-1} v ^i)$
and
\be
T^{(\rm m)}{}^0_{\ 0}  =  - \rho_{\rm m} \equiv  - \bar \rho_{\rm m} ( 1 + \deltam ) \;, \qquad T^{(\rm m)}{}^0_{\ i}  =  \rho_{\rm m} a   v ^i = - a^2 T^{(\rm m)}{}^i_{\ 0}\;, \qquad
T^{(\rm m)}{}^i_{\ j}  = \rho_{\rm m} v^i v^j\label{se}\;,
\ee
where we have defined $\delta $ and $ v^i$, respectively  the  energy density contrast and 3-velocity of matter, and $\bar \rho_{\rm m}(t)$ is the background energy density.  Importantly, though, in the mildly nonlinear regime, dark matter is not a perfect collisionless fluid.  Indeed, describing the deviations from this  behavior is the aim of the EFTofLSS, and will be discussed thoroughly in the rest of this paper.

In this paper, we work with scalar perturbations to the metric in Newtonian gauge, which can be written
\be
\label{metric_Newtonian}
ds^2 = - (1+2 \Phi) dt^2  + a^2(t) (1-2 \Psi)  \delta_{ij} d x^i d x^j \;.
\ee
We also work in the quasi-static, non-relativistic limit.  A convenient way to keep track of the terms that are important in this limit is to introduce a small counting parameter $\epsilon$ such that a spatial derivative counts as $\partial_i/Ha \sim {\cal O}(\epsilon^{-1})$. 
{For instance, relevant for the fluid-like equations, we have 
\be
\label{epsilon}
\Phi \sim \epsilon^2 \;, \qquad v^i  \sim \frac{\partial_i \Phi}{Ha}   \sim \epsilon \;, \qquad \delta \sim \frac{\partial^2 \Phi}{(Ha)^2}   \sim 1\; ,
\ee
where $\partial^2 = \sum_{i=1}^3 \partial_i \partial_i$. Time derivatives are of order ${\cal O} (H)$ and thus  do not change the order of $\epsilon$, e.g.~$\dot \Phi \sim  {\cal O} (\epsilon^2 H) $.
The leading operators in this expansion are those  with the highest number of spatial derivatives per number of fields. 
For  operators with $n$  fields, the dominant ones  contain $2(n-1)$ spatial derivatives and are thus of order $\epsilon^2$ \cite{CLV1}. 
The others are of order ${\cal O} (\epsilon^4)$ and contribute to post-Newtonian corrections. 

Defining the three vector 
\be \label{phiadef}
\varphi_a\equiv\left(
\begin{array}{c}
\Phi\\
\Psi\\
\chi
\end{array}
\right) \ ,
\ee
to write equations in a compact form, at leading order in the above expansion, the  Newtonian gauge actions at quadratic, cubic, and quartic order in  perturbations are respectively given by  \cite{CLV1}
\be
\begin{split}
\label{total_action_pert_c}
S_{\rm g}^{(2)}   &= - \int d^3 x dt \,   a \, M^2  A_{ab}(t) \partial_i \varphi_a \partial_i \varphi_b  \;, \\
S_{\rm g}^{(3)}  &= \int d^3 x dt \,  \frac{M^2}{ 3!\,  a H^2} B_{abc} (t)\varepsilon^{ikm} \varepsilon^{jl m} \partial_i  \varphi_a \partial_j \varphi_b \partial_k \partial_l \varphi_c  \;,  \\
S_{\rm g}^{(4)} &=  \int d^3 x dt \, \frac{M^2}{ 4! \, a^3 H^4} C_{abcd} (t)\varepsilon^{ikm} \varepsilon^{jl n} \partial_i \varphi_a \partial_j \varphi_b  \partial_k \partial_l \varphi_c \partial_m \partial_n \varphi_d  \;,
\end{split}
\ee
where the dimensionless arrays $A_{ab}$, $B_{abc}$  and  $C_{abcd}$ parametrize the coupling strength between fields and $\varepsilon^{ijk}$ is the 3-dimensional Levi-Civita symbol. Their explicit definition is given in App.~\ref{app:definitions}.
Variation  of the action with respect to the fields yields the equations of motion,
\be\label{masterapp}
A_{da}\partial^2 \varphi_a  - \frac{B_{dab}}{4 H^2 a^2} \varepsilon^{ikm} \varepsilon^{jl m}  \partial_i \partial_j \varphi_a \partial_k \partial_l \varphi_b-\frac{C_{dabc}}{ 12 H^4 a^4}  \varepsilon^{ikm} \varepsilon^{jl n} \partial_i \partial_j \varphi_a  \partial_k \partial_l \varphi_b \partial_m \partial_n \varphi_c = \deltac_{d1} \,\frac{\bar \rho_{\rm m} a^2   }{2 M^2} \, \deltam \,.
\ee}

\section{The effective fluid  of modified gravity} 
\label{Effective fluid}

\subsection{Effective fluid equations with only dark matter} \label{dmonlysec}

As a warm-up, in this subsection we briefly review the construction of the effective fluid equations for dark matter \cite{Baumann:2010tm,Carrasco:2012cv} (for a more rigorous treatment dealing also with the higher moments of the Boltzmann hierarchy see \cite{Baumann:2010tm}).  The general approach is to derive the equations for the full fields, which includes both long wavelength and short wavelength perturbations, and then smooth them to obtain the equations for the long-wavelength parts of the fields.  This procedure will introduce an effective stress tensor which describes the effects of short scale physics on the long modes.  Throughout this subsection, we will assume that $\Phi = \Psi$, up to relativistic corrections.  First, we expand the Einstein tensor into a background part $\bar G^\mu{}_\nu$, a linear part $G_{\rm L}{}^\mu{}_\nu $ and a remaining nonlinear part $G_{\rm NL}{}^\mu{}_\nu$, i.e., 
\be
G^\mu{}_\nu = \bar G^\mu{}_\nu +G_{\rm L}{}^\mu{}_\nu + G_{\rm NL}{}^\mu{}_\nu \;.
\ee
 Then, the perturbed Einstein equations are  
\be
\label{eins1}
M_{\rm Pl}^2 \big( G_{\rm L}{}^\mu{}_\nu + G_{\rm NL}{}^\mu{}_\nu \big) = \delta T^{\rm (m)}{}^\mu{}_\nu \;,
\ee
where  $\delta T^{\rm (m)}{}^\mu{}_\nu$ contains all pieces of the stress tensor besides the background.  Following \cite{Baumann:2010tm}, we start with the linearized Bianchi identity
\be \label{linearbian}
\bar \nabla_\mu G_{\rm L}{}^\mu{}_\nu + \nabla^{\rm L}_\mu \bar G^\mu{}_\nu = 0  \ , 
\ee
where quantities with an over-bar are evaluated on the background, and quantities with an $\rm L$ are linear in perturbations.  

We start by evaluating this for $\nu = 0$.  After using the Einstein equations \eqref{eins1} to plug in the expression for $G_{\rm L}{}^{\mu}{}_0$, and dropping relativistic corrections, we find
\be \label{conteq111}
a^{-3} \partial_\mu \left( a^3 \delta T^{(\rm m ) }{}^\mu{}_0  \right) = 0 \ .
\ee
Now, using the matter stress tensor in the form of a fluid as in \eqn{matterstresstensor} and combining \eqn{conteq111} with the zeroth order equation for $\bar \rho_{\rm m}$, i.e. $\dot{ \bar \rho}_{\rm m}  = - 3 H \bar \rho_{\rm m} $, we obtain the standard continuity equation in the non-relativistic limit, 
\be \label{anothercont}
  \dotrhom  + 3 H \rhom + a^{-1} \partial_i \left( \rhom v^i \right) = 0 \ . 
\ee

Now we move on to the Euler equation.  After using that $\bar \Gamma^\sigma_{\mu i} G_{\rm L}{}^\mu{}_\sigma = 0$, the linear Bianchi identity \eqref{linearbian} for $\nu = i$ becomes 
\be \label{lineareqbianchi}
a^{-3} \partial_\mu \left( a^3 G_{L}{}^\mu{}_i \right) = 2 \dot H \partial_i \Phi \ .
\ee
Again, we use \eqn{eins1} to replace $G_{\rm L}{}^{\mu}{}_i$ and get
\be \label{intereuler}
a^{-3} \partial_\mu \left( a^3 \delta T^{(\rm m ) }{}^\mu{}_i  \right) = a^{-3} \mpl^2 \partial_\mu \left( a^3   G_{\rm NL}{}^\mu{}_i \right)  - \bar \rho_{\rm m} \partial_i \Phi \ ,
\ee
where on the right-hand side we have used that $ 2\dot H \mpl^2 = - \bar \rho_{\rm m}$.
At leading order in $\epsilon$, the nonlinear part of the Einstein tensor reads
\be
\label{GNLexp}
G_{\rm NL}{}^j{}_i = a^{-2} \left(   \delta^{j}{}_i (\partial_k \Phi)^2 -2  \partial_i \Phi \partial_j \Phi  \right)   \ ,
\ee
which in \eqn{intereuler} contributes a term of the same order as the other two.  One can also check that $ \partial_0 G_{\rm NL}{}^0{}_i $ is down by ${\cal O}( \epsilon^2) $ with respect to $ \partial_j G_{\rm NL}{}^j{}_i $, so that we can ignore it in \eqn{intereuler}.
Thus, we are left with 
\be
\begin{split}
a^{-3} \partial_\mu \left( a^3 \delta T^{(\rm m ) }{}^\mu{}_i  \right)  & =  - \partial_j \left[ \bar \rho_{\rm m} \Phi  \delta^{j}{}_i -  a^{-2} \mpl^2 \left(    \delta^{j}{}_i (\partial_k \Phi)^2 -2  \partial_i \Phi \partial_j \Phi  \right)  \right]  \label{eqnnumber1}  \\
& =  - \rhom \, \partial_i \Phi   
\ ,
\end{split}
\ee
where to get the second equality we have used  the Poisson equation, $2 M_{\rm Pl}^2 a^{-2} \partial^2 \Phi = \bar \rho_{\rm m} \delta$.  

We have written the right-hand side of the above equation in two different forms to stress two different points about this equation.  First, using the right-hand side in the second form, replacing the expression for the stress tensor \eqref{matterstresstensor} on the left-hand side and using the continuity equation \eqref{anothercont}, \eqn{eqnnumber1} becomes the standard Euler equation in the non-relativistic limit, i.e.,
\be \label{anothereulereq}
 \rho_{\rm m}\left( \dot v^i + H v^i + \frac1{a} v^j \partial_j v^i \right) =  - \frac{1}{a} { \rho_{\rm m}}  \partial_i \Phi   \;.
\ee
The second point concerns the smoothing procedure, which we briefly review next.

As described in \cite{Baumann:2010tm,Carrasco:2012cv}, one can define the long-wavelength fields by smoothing with a window function $W_\Lambda ( \xvec - \xvec')$. In particular, let us define the long-wavelength gravitational potential $\Phi_{\ell}$, the long-wavelength dark matter density contrast $\delta_{\ell}$ and momentum density $\pi^i_\ell \equiv \bar \rho_{\rm m}  (1+\delta_{\ell})  v_{\ell}^i  $ respectively as 
\begin{align}
\Phi_{\ell} ( \xvec )  & \equiv \int d^3 x' \, W_\Lambda ( \xvec - \xvec') \, \Phi(\xvec') \;, \\
\delta_{\ell} ( \xvec )  & \equiv \int d^3 x' \, W_\Lambda ( \xvec - \xvec') \, \delta(\xvec') \;,  \\
\left(1+\delta_{\ell} ( \xvec ) \right) v_{\ell}^i ( \xvec ) & \equiv \int d^3 x' \, W_\Lambda ( \xvec - \xvec') \label{vlong} \, \left(1+ \delta(\xvec') \right) v^i ( \xvec  ' ) \ . 
\end{align}
Then, we apply this window function to the continuity equation \eqref{anothercont} and the Euler equation \eqref{anothereulereq}.  
Because we have chosen to smooth directly the momentum density in \eqn{vlong}, the smoothing of the continuity equation \eqn{anothercont} is simple and gives 
\be \label{deltacont}
\dot \delta_\ell + a^{-1} \partial_i \left( (1+\delta_\ell) v^i_{\ell} \right) = 0 \ .
\ee
The fact that no counterterms show up in \eqn{deltacont} means that the density $\delta_\ell$ and the momentum density $\pi^i_\ell $ are renormalized simultaneously.  However, because $v^i_\ell ( \xvec ) = \pi^i_\ell ( \xvec) /( \bar \rho_{\rm m} (1 + \delta_\ell (\xvec)))  $ is a contact operator, the velocity field itself must be renormalized \cite{Carrasco:2013mua}.\footnote{Alternatively, one can decide not to renormalize the velocity field directly, but instead add counterterms to \eqn{deltacont} \cite{Mercolli:2013bsa}.  We adopt the former approach in this paper, although, since we will not be considering correlations of the velocity field, this choice is of no consequence here.}

The situation is different for the Euler equation because of the presence of terms that mix long and short modes.  Reference \cite{Baumann:2010tm} describes the smoothing of the Euler equation in great detail, so here we simply report the result and highlight one important point.  Applying the smoothing procedure to \eqn{eqnnumber1}, we find that 
\be \label{anothereulereqsmooth}
 \rho_{\rm m , \ell}\left( \dot v_\ell^i + H v_\ell^i + \frac1{a} v_\ell^j \partial_j v_\ell^i   + \frac{1}{a}  \partial_i \Phi_\ell  \right) = -  \frac{1}{a} \partial_j \tau^{ij}_s  \;,
\ee
where $\tau^{ij}_s$ is made up of all short modes and  describes how the short modes affect the dynamics of the long modes.  Because the short modes are not accessible in perturbation theory, this is an incalculable object.  To deal with this, we parametrize our ignorance by expanding $\tau^{ij}_s$ in powers and derivatives of the long wavelength fields, and we include all operators, called counterterms, that are consistent with the equivalence principle.  The important thing about the structure of the right-hand side of \eqn{anothereulereqsmooth} is that it is a total derivative, and this is ensured by the form of the first equality of  \eqn{eqnnumber1}.  This structure dictates the leading $k$ dependence of the counterterms, as we will see in Sec.~\ref{countertermsec}.

The final form of the continuity, Euler, and Poisson equations, then, is
\begin{align}  \label{finalconteq}
& \dot \delta_\ell + a^{-1} \partial_i \left( (1+ \delta_\ell ) v^i_{\ell} \right) = 0 \ , \\
&  \dot v_\ell^i + H v_\ell^i + \frac1{a} v_\ell^j \partial_j v_\ell^i   + \frac{1}{a}  \partial_i \Phi_\ell   = -\frac{1}{a \, \rho_{\rm m , \ell}}  \partial_j \tau^{ij}_s \ ,    \label{finaleulereq}  \\
& \partial^2 \Phi_\ell = \frac{3 }{2} H^2  a^2 \,   \omegam \delta_\ell \  \ , \label{finalPoissoneq}
\end{align}
where $\Omega_{\rm m} \equiv \bar \rho_{\rm m} /(3 M_{\rm Pl}^2 H^2)$.

\subsection{Effective fluid equations with dark energy} \label{defluid}

Now, we would like to extend this analysis by including dark energy and modified gravity in the quasi-static limit, using the EFTofDE action \eqref{total_actionEFT}.\footnote{F.V.~is in debt with L.~Alberte, P.~Creminelli and J.~Gleyzes for many interesting conversations about the subject of this section.}  Because matter and dark energy are not directly coupled in the Jordan frame, the continuity equation \eqref{finalconteq} is not changed, so we will focus on the Euler equation. In particular, the  goal is to find that the short modes enter the Euler equation as a total derivative, similarly to what happens in the pure dark matter case, see \eqn{anothereulereqsmooth}.  The reason that this is not obvious anymore is because the Poisson equation, which we had to use  in Sec.~\ref{dmonlysec},  is no longer linear in $\delta$ but it is now modified. Indeed, in the presence of modifications of gravity 
this equation is replaced by a more complicated system of equations \eqn{masterapp} whose perturbative solution at third order in perturbations is given, for $\partial^2 \Phi$, by \eqn{sol_NL1} below.

The full action that we are considering is $S = S_{ \rm g} + S_{\rm m}$ where $S_{\rm g}$ is the gravitational action \eqref{total_actionEFT}, which depends on the metric and the dark energy field $\chi$; $S_{\rm m}$ is the dark-matter action, whose stress tensor is given by \eqn{matterstresstensor}.  
To proceed, we define the gravitational tensor $T^{\rm  (g)}{}_{\mu \nu}$ from the variation of the gravitational action with respect to the metric,
\be
T^{\rm  (g)}{}_{\mu \nu} \equiv - \frac{2}{\sqrt{-g} } \frac{\delta S_{\rm g}}{\delta g^{\mu \nu}} \; .
\ee
(In the absence of dark energy and modified gravity this is simply $- M_{\rm Pl}^2 G_{\mu \nu}$.)
Thus, including matter, the perturbed equations for the metric read
\be
\label{Einsteineqs}
\delta T^{\rm  (g)}{}_{\mu \nu} +  \delta T^{\rm (m)} {}_{\mu \nu}=0\;.
\ee
Next, let us split $\delta T^{\rm (g)}{}_{\mu \nu}$ into a linear and nonlinear part,
\be
\delta T^{\rm (g)}{}_{\mu \nu}\equiv  T^{\rm (g)}_{\rm L}{}_{\mu \nu}+  T^{\rm (g)}_{\rm NL}{}_{\mu \nu} \;.
\ee
Analogous to \eqn{lineareqbianchi}, we have the following identity at linear order,\footnote{While the Bianchi identity is no longer valid once we break diffeomorphism invariance, there is still a linear version coming from spatial diffeomorphisms.  Consider the diffeomorphism $x^\mu \rightarrow x^\mu + \xi^\mu ( x )$, but with $\xi^0 = 0$.   This induces $\delta g^{\mu \nu} = \nabla^\mu \xi^\nu + \nabla^\nu \xi^\mu$ and $ \delta \pi = - \xi^\mu \partial_\mu \pi$, which gives a variation in the gravitational action of 
\be \label{lavendar}
\delta S_{\rm g} = - \int d^4 x \sqrt{-g} \left(   T^{(\rm g)}_{\mu \nu} \nabla^\mu \xi^\nu + \frac{1}{\sqrt{-g}} \frac{\delta S_{\rm g}}{\delta \pi} \xi^\mu \partial_\mu \pi \ \right) \ . 
\ee
Now, because the action starts at quadratic order, $\delta S_{\rm g } / \delta \pi$ starts at first order, so that the second term above starts at second order in perturbations.  This means that at linear order the first term must be zero itself, which gives 
\be
\bar \nabla_\mu T^{( \rm g)}_{\rm L} {}^\mu{}_i + \nabla^{\rm L}_\mu \bar T^{(\rm g)} {}^\mu{}_i = 0 \ .
\ee
Then, to get \eqn{apples}, one follows the same steps that lead to \eqn{lineareqbianchi}.
 } 
\be \label{apples}
a^{-3} \partial_\mu \left(a^3 T^{\rm (g)}_{\rm L}{}^\mu{}_i \right) =  \frac{\bar \rho_{\rm m}}{M^2} \partial_i \Phi  \;.
\ee
Thus, from eq.~\eqref{Einsteineqs} it follows that, on the equations of motion,
\be \label{deeulereq1}
a^{-3} \partial_\mu \left[ a^3 \left( \delta T^{\rm (m)}{}^\mu{}_i + T^{\rm (g)}_{\rm NL}{}^\mu {}_i \right) \right] =  - \frac{\bar  \rho_{\rm m}}{M^2} \partial_i \Phi  \;.
\ee

We want to derive an equation analgous to \eqn{eqnnumber1} from this equation.  As in \eqref{eqnnumber1}, the leading terms coming from the matter stress tensor enter \eqn{deeulereq1} at order $ \epsilon$, so we need to examine the leading terms coming from the nonlinear gravitational stress tensor.  First, let us examine $\partial_j T^{\rm (g)}_{\rm NL}{}^{j}{}_{i}$.  By direct calculation (see the explicit expression in App.~\ref{stresstensorsec}, \eqn{TNLji}), one can check that $T^{\rm (g)}_{\rm NL}{}^{j}{}_{i}$ starts at $ {\cal O}(\epsilon^0)$, one order higher than the matter stress-tensor in the above equation.
However, one can check that the divergence of this term vanishes identically, so that there are no contributions to the Euler equation at this order of spatial derivatives, as expected.  Then we have to go to an order higher in $\epsilon$.
One can check that there is no contribution at ${\cal O}(\epsilon)$ to $T^{\rm (g)}_{\rm NL}{}^{j}{}_{i}$. At ${\cal O}(\epsilon^2)$ the expression for  $T^{\rm (g)}_{\rm NL}{}^{j}{}_{i}$  is long, but we do not need it  explicitly for our purposes, because this piece automatically enters as a total derivative at the correct order in $\epsilon$.  This leaves us with
\be \label{deeulereq2}
a^{-3} \partial_\mu \left( a^3 \delta  T^{\rm (m)}{}^\mu{}_i  \right) + a^{-3} \partial_0 \left( a^3 T^{\rm (g)}_{\rm NL}{}^0{}_i  \right) =  -  \partial_j \left( \frac{\bar  \rho_{\rm m}}{M^2} \delta^{j}{}_i \Phi   + T^{\rm (g)}_{\rm NL}{}^j {}_i \right)\; .
\ee

The final term that we have to consider is $T^{\rm (g)}_{\rm NL}{}^0{}_i$. By direct inspection (see again App.~\ref{stresstensorsec} for the explicit expression) 
it can be checked that at ${\cal O}( \epsilon)$ this term is a total derivative.
This means that it enters the Euler equation as a total derivative as well, so that the equation above can be rewritten as
\be \label{deeulereq3}
a^{-3} \partial_\mu \left( a^3 \delta T^{\rm (m)}{}^\mu{}_i  \right)  = - \partial_j \left(  \frac{\bar  \rho_{\rm m}}{M^2}  \delta_{i}^{j} \Phi   +  T^{\rm (g)}_{\rm NL}{}^j {}_i   + a^{-3} \partial_0 \left( a^3 t^{j}{}_i \right)  \right) \; ,
\ee
where we have defined the tensor $t^{j}{}_i$ as
\be
\label{TNL0i}
\begin{split}
  \partial_j t^{j}{}_i \equiv T^{\rm (g)}_{\rm NL}{}^{0}{}_{i} \;.
\end{split}
\ee
Moreover, it is very lengthy but straightforward to check that, by plugging the explicit expressions for $T^{\rm (g)}_{\rm NL}{}^j {}_i$ and $t^{j}{}_i$ at the relevant order in $\epsilon$ on the right-hand side of eq.~\eqref{deeulereq3} and using the nonlinear {constraint equations  \eqref{masterapp},} 
eq.~\eqref{deeulereq3} can be also written as 
\be \label{deeulereq2tot}
a^{-3} \partial_\mu \left( a^3 \delta T^{\rm (m)}{}^\mu{}_i  \right)  =  - \frac{  \rho_{\rm m}}{M^2} \partial_i   \Phi  \; .
\ee

In summary, eq.~\eqref{deeulereq3} shows explicitly that in the quasi-static limit any corrections to the Euler equation due to short-distance physics must enter as the divergence of an effective stress tensor.  
As the final step, we use the expression for the matter stress tensor \eqn{matterstresstensor} and smooth \eqref{deeulereq3} to obtain the fluid-like equations.  
Since the continuity equation is the same as in \eqn{finalconteq}, the smoothing of the Euler equation \eqref{deeulereq2tot} gives the final form of the fluid-like equations for the long-wavelength fields as 
\begin{align}  \label{finalconteq1}
& \dot \delta + a^{-1} \partial_i \left( (1+\delta) \, v^i \right) = 0 \ , \\
&  \dot v^i + H v^i + \frac1{a} v^j \partial_j v^i   + \frac{1}{a}  \partial_i \Phi   = -\frac{1}{a \, \rho_{\rm m }}  \partial_j \tau^{ij}_s    \label{finaleulereq1}  \   ,
\end{align}
where here we have dropped the subscript $_\ell$ to remove clutter. 

Note that the situation that we discuss here is different from the case of clustering quintessence for small sound speed \cite{Creminelli:2009mu,Sefusatti:2011cm,Lewandowski:2016yce}, where the dark energy behaves as a second dynamical fluid. More generally,  multiple fluids  (see e.g.~\cite{Lewandowski:2014rca,Lewandowski:2016yce,Senatore:2017hyk}), 
 can exchange momentum between themselves through their interactions with gravity. As a consequence, counterterms could enter the Euler equation through  an effective force term $\gamma^i_s (\xvec , a)$ that is not a total derivative \cite{Lewandowski:2014rca}. The quasi-static assumption made in this article ensures that the dark energy field satisfies constraint equations that can be used to re-express the scalar field fluctuations in terms of matter fluctuations and that there is no separate independent dark energy fluid.

Coming back to eq.~\eqref{finaleulereq1}, to get the expression for $\partial^2 \Phi_\ell$, the equivalent of the Poisson equation \eqref{finalPoissoneq} in the case of pure dark matter,
we must perturbatively solve the {system of constraint equations \eqref{masterapp} which arises from  varying the gravitational and matter actions; see   \cite{CLV1} for details.} 
While the field $\chi$ is already an effective long-wavelength field, the metric potentials $\Phi$ and $\Psi$ are the full fields, and so these equations must be smoothed to be written in terms of the long-wavelength fields.  However, one can check that the constraint equations  \eqref{masterapp} are linear in $\Phi$ and $\Psi$, and thus the smoothing can be done without adding additional counterterms.  Dropping the subscript $_\ell$ to reduce clutter,  {to order $\delta^3$ one finds \cite{CLV1}} 
 \begin{align}
\label{sol_NL1}
\partial^2 \Phi= & \ H^2 a^2 \bigg\{ \frac{3 \, \omegam }{2}  \,  \mu_{\Phi } \,  \deltam+  \left( \frac{3\, \omegam}{2}  \right)^2 \mu_{\Phi,2}  \left[\deltam^2-\left(\partial^{-2}{\partial_i\partial_j}\deltam\right)^2\right]  \\
&+\left(  \frac{3\, \omegam}{2} \right)^3 \mu_{\Phi,22}\left[\deltam-\left(\partial^{-2} {\partial_i\partial_j} \deltam\right) \partial^{-2} {\partial_i\partial_j} \right]\left[\deltam^2-\left(\partial^{-2}{\partial_k\partial_l} \deltam\right)^2\right]   \nonumber  \\
&+\left(  \frac{3\, \omegam}{2}  \right)^3 \mu_{\Phi,3}\left[\deltam^3-3\deltam\left(\partial^{-2}{\partial_i\partial_j} \deltam\right)^2+2 (\partial^{-2}{\partial_i\partial_j} \deltam)( \partial^{-2}{\partial_k\partial_j}\deltam )( \partial^{-2}{\partial_i\partial_k}\deltam )\right] \bigg\}   + {\cal O} (\delta^4)\,, \nonumber
\end{align}
where 
\be
\Omega_{\rm m} \equiv \frac{\bar \rho_{\rm m} }{3 M^2 H^2} \; .
\ee
{The functions $\mu_{\Phi} ( a )$, $\mu_{\Phi,2} ( a )$, $\mu_{\Phi,22} ( a )$, and $\mu_{\Phi , 3 }(a)$ are related to the coefficients of the action \eqn{total_actionEFT} \cite{CLV1}.  Their expressions are explicitely given in App.~\ref{app:definitions} but from the viewpoint of the LSS equations, they are simply free functions of time.}
We stress that this solution is only valid on scales above the nonlinear scale where $\delta \sim 1$ and above the Vainshtein scale where scalar field fluctuations enter the nonlinear regime, as shown in our companion article \cite{CLV1}.

As the last piece to the puzzle, we will give the explicit expansion, in terms of the long-wavelength fields, of the effective stress tensor appearing in \eqn{finaleulereq1} in Sec.~\ref{Perturbation theory} when we discuss the perturbative solution.

%
%
%
%
%

%
%
%
%
%

\section{Calculation of the one-loop power spectrum}\label{Perturbation theory}

In this section, we solve \eqn{finalconteq1} - \eqn{sol_NL1} for the one-loop power spectrum of dark matter density fluctuations in the presence of the dark-energy operators presented above.  For the one-loop computation that concerns us here, we need to solve for $\delta$ up to third order, including the counterterm contribution from the EFTofLSS.

\subsection{Equations in Fourier space}

{Let us define the conformal Hubble rate as $\cH \equiv Ha$ and use a prime to denote the derivative with respect to the scale factor $a$.}
In Fourier space, and in terms of the scale factor $a$, the equations of motion for the dark-matter overdensity $\delta$ and the rescaled velocity divergence,
\be
\Theta \equiv - \partial_i v^i / \cH \;,  
\ee
are 
\begin{align} 
a \, \delta ' ( \kvec , a ) - \Theta ( \kvec , a ) & =  \int_{\kvec_1} \int_{\kvec_2}  ( 2 \pi )^3 \delta_D ( \kvec - \kvec_1 - \kvec_2 ) \nonumber  \\
& \hspace{0.5in} \times \alpha ( \kvec_1 , \kvec_2 ) \Theta ( \kvec_1 , a ) \delta ( \kvec_2 , a) \label{conteq1} \\
a \, \Theta ' ( \kvec , a ) + \left( 1 + \frac{ a \cH'}{\cH} \right) \Theta ( \kvec , a ) + \frac{k^2}{\cH^2}  \Phi ( \kvec , a)  & =  \int_{\kvec_1} \int_{\kvec_2}  ( 2 \pi )^3 \delta_D ( \kvec - \kvec_1 - \kvec_2 ) \nonumber   \\
& \hspace{0.5in} \times \beta ( \kvec_1 , \kvec_2 ) \Theta ( \kvec_1 , a ) \Theta ( \kvec_2 , a)\nonumber  \\
& \hspace{0in} + \cH^{-2} \int d^3 x\, e^{i \kvec \cdot \xvec}  \partial_i \left( \rho_{\rm m}^{-1}\partial_j \tau^{ij}_s (\xvec , a) \right)   \label{eulereq1}
\end{align}
where  $\alpha$ and $\beta$ are the standard dark matter interaction vertices,
\begin{align}  
\alpha ( \qvec_1 , \qvec_2 ) = 1 + \frac{\qvec_1 \cdot \qvec_2}{q_1^2}  \hspace{.3in} \text{and} \hspace{.3in}
\beta( \qvec_1 , \qvec_2 )  = \frac{ | \qvec_1 + \qvec_2 |^2 \qvec_1 \cdot \qvec_2}{2 q_1^2 q_2^2}    \label{betadef2new} \ ,
 \end{align}
 and we have used the notation $\int_{\kvec} \equiv \int \frac{ d^3 k}{( 2 \pi )^3}$.
 
As discussed above, because dark matter and dark energy are coupled only through gravity (in the Jordan frame), the above equations are exactly the same as in the dark-matter-only case.  The modification of gravity comes through a modified relation between $\partial^2 \Phi$ and $\delta$, i.e.~\eqn{sol_NL1}.
{In Fourier space, \eqn{sol_NL1} reads,}\footnote{The analogous equations for $\partial^2 \Psi$ and $\partial^2 \chi$ in terms of $\delta$ are given in  \cite{CLV1}.  Once we solve for the one-loop power spectrum of $\delta$ in the rest of this work, this expression of $\partial^2 \Phi$ (and the analogous ones for $\partial^2 \Psi$ and $\partial^2 \chi$) can be used to straightforwardly compute the one-loop correlation functions of the potentials.  Having these expressions is useful for computing other important observables used to test dark energy, e.g.~the total lensing potential $( \Phi + \Psi )/2$.  } 
\begin{align} \label{newd2phi}
- \frac{ k^2}{\cH^2} \Phi ( \kvec,a ) & =  \mu_{\Phi }  \frac{3 \, \omegam   }{2 } \, \delta ( \kvec,a  ) \\
& \hspace{-.3in} + \mu_{\Phi , 2}   \left( \frac{ 3 \, \omegam}{2  } \right)^2     \int_{\kvec_1} \int_{\kvec_2} ( 2 \pi )^3 \delta_D ( \kvec - \kvec_1 - \kvec_2 ) \,  \gamma_2 ( \kvec_1 , \kvec_2 ) \delta ( \kvec_1,a ) \delta( \kvec_2 ,a) \nonumber \\
& \hspace{-.3in} +  \mu_{\Phi , 3}   \left( \frac{ 3 \, \omegam  }{2  } \right)^3   \int_{\kvec_1} \int_{\kvec_2} \int_{\kvec_3}  ( 2 \pi)^3 \delta_D ( \kvec - \kvec_1 - \kvec_2 - \kvec_3) \gamma_3 ( \kvec_1 , \kvec_2 , \kvec_3) \delta ( \kvec_1 ,a) \delta( \kvec_2,a ) \delta ( \kvec_3  ,a) \nonumber \\
& \hspace{-.3in} +   \mu_{\Phi , 22}   \left( \frac{ 3 \, \omegam }{2  } \right)^3    \int_{\kvec_1} \int_{\kvec_2} \int_{\qvec_1} \int_{\qvec_2}  ( 2 \pi)^3 \delta_D ( \kvec - \kvec_1 - \kvec_2) ( 2 \pi)^3 \delta_D ( \kvec_2 - \qvec_1 - \qvec_2) \nonumber \\
& \hspace{1.4in} \times \gamma_2 ( \kvec_1 , \kvec_2  ) \gamma_2 ( \qvec_1 , \qvec_2) \delta ( \kvec_1 ,a ) \delta( \qvec_1 ,a) \delta ( \qvec_2  ,a) \nonumber \ , 
\end{align}
where  the momentum dependent interaction vertices describing the effects of dark energy are given by
\be
\label{gammas}
\begin{split}
\gamma_2 ( \kvec_1 , \kvec_2 ) & = 1 - \frac{ \big( \kvec_1 \cdot \kvec_2 \big)^2}{k_1^2 k_2^2}\\
\gamma_3 ( \kvec_1 , \kvec_2 , \kvec_3 ) & = \frac{1}{k_1^2 k_2^2 k_3^2 } \Big( k_1^2 k_2^2 k_3^2  + 2 \big( \kvec_1 \cdot \kvec_2\big)  \, \big( \kvec_1 \cdot \kvec_3 \big) \, \big( \kvec_2 \cdot \kvec_3 \big)     \\
& \hspace{1in} - \big( \kvec_1 \cdot \kvec_3 \big)^2 k_2^2 -  \big( \kvec_2 \cdot \kvec_3 \big)^2  k_1^2 - \big( \kvec_1 \cdot \kvec_2 \big)^2 k_3^2 \Big) \ . 
\end{split}
\ee
Similar results have been found in the context of Horndeski theories \cite{Takushima:2013foa,Takushima:2015iha,Bose:2016qun}.  The linear equations are modified by the term proportional to $\mu_{\Phi}$ and new nonlinear terms are introduced by the new nonlinear vertices of this equation.  The vertex proportional to $\mu_{\Phi,3}$, which is truly a cubic vertex (i.e.~it is not built out of two quadratic vertices), is of a new form in large-scale structure.  We will comment more specifically on the effects of these vertices later.  
Finally, as mentioned in Sec.~\ref{defluid}, the effective stress tensor $\tau^{ij}_s$ is sourced by long-wavelength fluctuations and we will give its form in more detail in Sec.~\ref{countertermsec}.

%
%

\subsection{Solutions and Green's functions} \label{greenfunsec}

To solve the above equations, we seek a perturbative expansion of the dynamical fields in the form
\be \label{pertexp}
\delta( \kvec , a  ) = \sum_{n = 1 }^\infty \delta^{(n)} ( \kvec , a ) + \delta^{\rm ct} ( \kvec , a )  \hspace{.3in} \text{and} \hspace{.3in}  \Theta ( \kvec , a  ) = \sum_{n = 1 }^\infty \Theta^{(n)} ( \kvec , a ) + \Theta^{\rm ct} ( \kvec , a ) \;,
\ee
where $\deltan$ and $\thetan$ are the $n$-th order solutions to the equations of motion in the absence of the effective stress tensor $\tau^{ij}_s$, and $\delta^{\rm ct}$ and $\Theta^{\rm ct}$ are the fields sourced by $\tau^{ij}_s$.

To find the linear equations of motion, we combine the first line of the expression for $\partial^2 \Phi$ from \eqn{newd2phi} with the continuity and Euler equations from eqs.~\eqref{conteq1} and \eqref{eulereq1}.  This gives the linear equation of motion for $\delta^{(1)} ( \kvec , a )$,
\be  \label{lineaeq}
 a^2 \, \delta^{(1)}( \kvec , a ) '' + a \left( 2  + \frac{a \cH'(a)}{\cH(a)} \right)  \delta^{(1)}( \kvec , a ) '  - \mu_{\Phi}(a) \frac{ 3\, \omegam(a) }{2  } \delta^{(1)} ( \kvec , a ) = 0 \ .
\ee
This equation has two independent solutions. For small deviations from $\Lambda$CDM, one is a growing solution, denoted by $D_+(a)$, and the other is a decaying solution, denoted by $D_-(a)$.  
Thus, at late time the linear solution is given by the growing mode,
\be
\delta^{(1)} ( \kvec , a ) = \frac{D_+(a) }{D_+(a_i )} \delta^{(1)}( \kvec , a_i) \hspace{.4in} \text{and} \hspace{.4in} \Theta^{(1)} ( \kvec , a ) = \frac{ a D_+(a)'}{D_+(a_i)} \delta^{(1)} ( \kvec , a_i )  \ , \label{lin_sol}
\ee
where $a_i$ is the initial time at which we choose to set the initial conditions.  This should be early enough so that the system is still linear, but past radiation domination, so that our equations for $\delta$ are correct.

With the two linear solutions $D_+$ and $D_-$, we can construct the four Green's functions for the system \eqref{conteq1} and \eqref{eulereq1},  $G^\delta_1 (a, \tilde a)$, $G^\delta_2(a, \tilde a)$, $G^\Theta_1(a, \tilde a)$ and $G^\Theta_2(a, \tilde a)$. These are explicitly derived in App.~\ref{gfsec}. The Green's function $G^\delta_1$ encodes the response of $\delta$ to a perturbation to the continuity equation, $G^\delta_2$ encodes the response of $\delta$ to a perturbation to the Euler equation, and similarly for $\Theta$.
Then, the perturbative solutions    of the system can  be written as
 \begin{align}  \label{dtGreen}
 &\delta^{(n)} ( \vk , a) =\int^a_0 d\ta \bigg(G^{\delta}_{1}(a,\ta)S^{(n)}_1(\vk, \ta)+G^{\delta}_{2}(a,\ta)S^{(n)}_2(\vk, \ta)\bigg) \ ,\\
  &\Theta^{(n)} (\vk , a )=\int^a_0 d\ta \bigg(G^{\Theta}_{1}(a,\ta)S^{(n)}_1(\vk, \ta)+G^{\Theta}_{2}(a,\ta)S^{(n)}_2(\vk, \ta)\bigg) \ , \label{dtGreen2}
  \end{align}
where  the source terms $S^{(n)}_i$ are the $n$-th order expansion of the right-hand sides of eq.~\eqref{conteq1} for $i=1$, and eq.~\eqref{eulereq1} for $i=2$, after plugging in \eqn{newd2phi}.  In general, the $n$-th order source term is proportional to $n$ powers of the linear field, i.e. $S^{(n)} \sim [ \delta^{(1)}]^n$, and the $k$ dependence is dictated by the particular dependence of the nonlinear vertices.  For example, $S^{(2)} \sim [\delta^{(1) }]^2$, and $S^{(3)}$ contains two types of terms, the normal $\delta^{(1)} S^{(2)}$, and the new $[\delta^{(1)}]^3$ term.

%

\subsection{Power spectrum} \label{oneloopsec}

The power spectrum is defined as 
\be
\langle \delta( \kvec , a ) \delta(\kvec ' , a) \rangle = ( 2 \pi)^3 \delta_D ( \kvec + \kvec') P(k,a) \ . 
\ee
Then, using the perturbative expansion in \eqn{pertexp} and assuming Gaussian initial conditions,\footnote{The inclusion of primordial non-Gaussianities is straightforward \cite{Angulo:2015eqa, Assassi:2015jqa, Assassi:2015fma}.} we can expand the power spectrum  up to one-loop as
\be
P (k,a) = P_{11} ( k , a ) + P_{\text{ 1-loop}}(k,a) \;.
\ee
On the right-hand side,  $P_{11}$ is the linear contribution,
\be
P_{11} ( k , a )  =  \langle \delta^{(1)} ( \kvec , a )  \delta^{(1)} ( \kvec ' , a ) \rangle '    \; ,
\ee
where $\langle \cdots \rangle ' $ means that we have stripped the factor of $( 2 \pi)^3 \delta_D ( \kvec + \kvec')$, which must be present due to momentum conservation,  from the right-hand side.
Using eq.~\eqref{lin_sol}, the linear power spectrum is given by
\be
P_{11} ( k , a ) = \left( \frac{D_+(a)}{D_+(a_i)} \right)^2 P^{\rm in}_{k} \ , \qquad P^{\rm in}_{k} \equiv \langle \delta^{(1)} ( \kvec , a_i ) \delta^{(1)} ( \kvec' , a_i ) \rangle ' \;.
\ee 
The initial power spectrum $P^{\rm in}_{k}$ can be obtained from a linear Einstein-Boltzmann solver like CAMB \cite{Lewis:1999bs} or CLASS \cite{Lesgourgues:2011re}, if $\Lambda$CDM initial conditions are sufficient, or one of the recently developed Boltzmann codes that include linear effects of the EFTofDE \cite{Hu:2013twa,Bellini:2015xja,Huang:2012mt,Bellini:2017avd}.  

The one-loop contribution to the power spectrum is given as
\be
P_{\text{ 1-loop}}(k,a) \equiv  P_{22} ( k , a) + P_{13} ( k , a ) + P^{\rm ct}_{13} ( k , a )  \;,
\ee
where 
\be 
\label{psdefs}
\begin{split}
 P_{22} ( k , a ) &= \langle \delta^{(2)} ( \kvec , a )  \delta^{(2)} ( \kvec ' , a ) \rangle '   \; , \\ 
P_{13} ( k , a ) & = 2 \langle \delta^{(1)} ( \kvec , a )  \delta^{(3)} ( \kvec ' , a ) \rangle ' \; , \\
P^{\rm ct}_{13} ( k , a ) &= 2 \langle \delta^{(1)} ( \kvec , a )  \delta^{\rm ct} ( \kvec ' , a ) \rangle '  \, .
\end{split}
\ee
We will first move to the calculations of $P_{22}$ and $P_{13}$ and postpone the calculation of $P^{\rm ct}_{13}$ to the next subsection.

To compute $P_{22}$ and $P_{13}$ we need the second and third order  solutions in perturbation theory without the contribution of the effective stress tensor $\tau^{ij}_s$, respectively $\delta^{(2)}( \kvec , a )$ and $\delta^{(3)}( \kvec , a )$ in the expansion in eq.~\eqref{pertexp}. These can be obtained by using eq.~\eqref{dtGreen} and the explicit calculation is  given in App.~\ref{delta23}. 
From these solutions we obtain
\begin{align} \label{p22foryou}
P_{22} ( k, a) & = \int \momspmeas{q} \int_0^{a}  d a_2 \int_0^{a_2} d a_1 \, \,  p_{22}(a, a_1 , a_2 ; \kvec , \qvec) \ , \\ 
 P_{13} ( k, a) & = \int \momspmeas{q} \int_0^{a}  d a_2  \left(  p_{13}^{(1)} (a , a_2 ; \kvec , \qvec)  +   \int_0^{a_2} d a_1 \, \,  p_{13}^{(2)}(a, a_1 , a_2 ; \kvec , \qvec)  \right)  \ , \label{p13foryou}  
\end{align}
where the integrands of these expressions are given by
\be
\begin{split} \label{p22text2}
p_{22}(a, a_1 , a_2 ; \kvec , \qvec) & \equiv  \sum_{i = 1}^7 T^{(22)}_i (a, a_1 , a_2) F^{(22)}_i ( \kvec , \qvec) \ ,  \\
p_{13}^{(2)}(a, a_1 , a_2 ; \kvec , \qvec) & \equiv  \sum_{i = 1}^{10} T^{(13)}_i (a, a_1 , a_2) F^{(13)}_i ( \kvec , \qvec) \; ,\\
p_{13}^{(1)}(a , a_2 ; \kvec , \qvec) & \equiv  T^{(13)}_{11} (a , a_2) F^{(13)}_{11} ( \kvec , \qvec)  \;.  
\end{split}
\ee
The integrands are thus given as a sum over separable products of time-dependent and momentum-dependent contributions.\footnote{{References \cite{Takushima:2013foa,Takushima:2015iha}  have solved for the standard perturbation theory kernels of the one-loop power spectrum, with gravitational sector described by Horndeski theories.  They have shown that the number of independent terms is actually much smaller than  in \eqn{p22text2}, due to relationships among the momentum-dependent kernels and among the Green's functions.  
Because the numerical computation of the loops was not too demanding, we did not seek to simplify our expressions further. 
Moreover, in our presentation  the momentum integrals, which can be done independently of the dark-energy parameters, are computed separately from the time integrals, which one must compute for each set of dark-energy parameters.}} Their explicit expressions are reported in App. \ref{expexp}.

Let us make some comments here. In  $p_{22}$, the contributions to the sum for $i=1, \ldots,4$ come from the standard dark-matter vertices, which are functionally the same as the corresponding $\Lambda$CDM functions, but numerically different due to the modification of $D_+$ and $D_-$ by $\mu_{\Phi}$ in the linear equations of motion. The contributions to the sum for $i=5, 6,7$ are due to the new nonlinear terms coming from \eqn{newd2phi}, i.e.~they are proportional to $\mu_{\Phi,2}$ and $\mu_{\Phi,22}$. 
  In  $p_{13}$
 there are two types of terms:  $p_{13}^{(2)}$, which has two insertions of Green's functions, and $p_{13}^{(1)}$, which has one insertion of a Green's function. The latter comes from the new cubic vertex proportional to $\mu_{\Phi,22}$. The contributions to the sum for $i=1, \ldots,6$ come from the standard dark-matter vertices while those for $i=7, \dots ,11$ are due to the new nonlinear compulings.
In conclusion, with respect to the exact time dependence computation in $\Lambda$CDM, we have six additional momentum integrals and eight additional time coefficients to compute.  An interesting point to notice about the above expressions is that the cubic vertex proportional to $ \mu_{\Phi,3}$ does not contribute to the power spectrum at one loop.  As can be seen in App.~\ref{expexp}, this is because $\gamma_3 ( \kvec , \qvec, - \qvec) = 0$.  However, this vertex will contribute to the two-loop power spectrum, the one-loop bispectrum and the tree level trispectrum.  Because $\alpha_{\rm V3}$ only shows up in $\mu_{\Phi,3}$, this means that the new quartic vertex in the Horndeski Lagrangian (\ref{total_actionEFT}) does not contribute to the one-loop power spectrum.

As discussed in the Introduction, some of the kernels $F^{(22)}_i ( \kvec , \qvec)$ and $F^{(13)}_i ( \kvec , \qvec)$ in eq.~\eqref{p22text2} contain spurious IR divergences. However, the equivalence principle guarantees that these vanish once all the contributions are summed together in the equal-time one-loop power spectrum. This has been shown to be the case for the standard $\Lambda$CDM vertices. Since the modifications of gravity that we introduce do not violate the equivalence principle, this must be the case for the new contributions as well.  Similarly, the kernels $F^{(13)}_i ( \kvec , \qvec)$ contain spurious UV divergences that are known to vanish by mass and momentum conservation.  Because this remains the case in our setup, these spurious divergences  do not appear in the final result. 

When computing the loop integrands, it is convenient to formulate them in terms of  IR\&UV-safe versions, which automatically remove spurious divergences and ensure that these do not significantly affect the numerical computation.    We report the derivations and  expressions of the IR\&UV-safe versions of the one-loop integrals in App. \ref{irsafesec}.  We now turn to discuss the contribution from the short-scale stress tensor, $P_{13}^{\rm ct}$.

%
%
\subsection{Effective stress tensor and counterterms} \label{countertermsec}

Next, we move on to expressing the effective stress tensor in \eqn{eulereq1} in terms of the long-wavelength fields.   Because the equivalence principle is satisfied under our assumptions, the short modes can only be affected by tidal effects, so that the stress tensor can be written as an
expansion in derivatives and powers of the tidal fields, $\partial_i \partial_j \Phi$, $\partial_i \partial_j \Psi$ and $\partial_i \partial_j \chi$, and of the first derivative of the velocity $\partial_j v^i$, evaluated along the fluid flow.  Thus, at the lowest order in derivatives and fields\footnote{For example, we do not include a term like $\partial^2 v^i_{\rm m}$ because it is the same as $\partial^i \partial_j v^j_{\rm m}$ apart from vorticity, which is only generated at a higher order in perturbation theory.} the effective stress tensor takes the following form \cite{Carrasco:2013mua}
\be
\begin{split} 
\label{stresstensor10}
& -  \left( \frac{1}{\rho_{\rm m}} \partial_j \tau^{ij}_s \right) ( \xvec , a)  \\   
& \hspace{.5in} =    \int d a' \Big[   \kappa^{(\Phi)}( a , a' )\,    \partial^i \partial^2 \Phi ( \xvec_{\rm fl}( \xvec ; a,a'), a'  )   + \kappa^{(\Psi)}( a , a' )\,    \partial^i \partial^2 \Psi (  \xvec_{\rm fl}( \xvec ; a,a') , a' )    \\ 
&\hspace{1.1in}     + \kappa^{(\chi)}( a , a' )\,    \partial^i \partial^2 \chi ( \xvec_{\rm fl}( \xvec ; a,a') , a' )   +\kappa^{(v)}( a , a' )\, \frac{1}{H}\partial^i \partial_j v_{\rm m}^j (  \xvec_{\rm fl}( \xvec ; a,a')  , a')     \\ 
& \hspace{1.1in}   + \kappa^{(\rm stoch.)}_1 ( a , a' )  \Delta^i_{\rm stoch.} (\xvec_{\rm fl}( \xvec ; a,a') , a' )     + \ldots \Big ] \;,
\end{split}
\ee
where the fluid line element $\xvec_{\rm fl}$ is defined recursively by
\be \label{xfldef}
\xvec_{\rm fl} ( \xvec ; a , a' ) = \xvec - \int_{a'}^a \frac{d a'' }{H \left( a'' \right) } \vec{v}_{\rm m} \left( \xvec_{\rm fl} ( \xvec ; a , a'' ) , a'' \right) \ ,
\ee
and $ \Delta^i_{\rm stoch.}$ is a stochastic term \cite{Carrasco:2012cv}. The $\kappa^{(i)}$ are free functions whose explicit dependence  on $a$ and $a'$ is not relevant for what follows.
The   above espression of the stress tensor shows that the EFTofLSS is non-local in time \cite{Carrasco:2013mua} (see also \cite{Carroll:2013oxa, Senatore:2014eva}).
Because for the one-loop computation we only need to work at linear order in the counterterms, we can set $\xvec_{\rm fl } = \xvec$ in \eqn{stresstensor10}.  For higher order corrections due to the fluid flow, one should expand the arguments for $\xvec_{\rm fl} $ near $ \xvec$.

Let us now look at the various terms appearing in \eqn{stresstensor10}.  First, the stochastic term is expected to be Poisson-like and does not correlate with the matter fields, so that in momentum space we can write, at lowest order in $k/k_{\rm NL}$, 
\be \label{stochcor}
\langle  \Delta^i_{\rm stoch.} ( \kvec )  \Delta^j_{\rm stoch.} ( \kvec ') \rangle = \frac{ ( 2 \pi)^3}{\knl^3} \delta ( \kvec + \kvec ' ) \frac{k^i k^j}{\knl^2} C_{\rm stoch.}  \;,   
\ee
where $C_{\rm stoch.}$ is expected to be an order-one number.  To get the contribution to the power spectrum, we contract the above with $k_i k_j$, so that the overall contribution is proportional to $k^4$ and, as we will see, it can be neglected with respect to the other contributions.\footnote{As discussed in \cite{Carrasco:2012cv}, the correlation function \eqref{stochcor} starts at order $k^i k^j$ because of momentum conservation, i.e.~because dark matter does not exchange momentum with dark energy, $\gamma^i_s (  \xvec , a )   = 0$.  If  $\gamma^i_s ( \xvec,a )   \neq 0$, a lower order $k^2$ contribution  is possible  \cite{Lewandowski:2014rca}.  In any case, stochastic contributions are expected to be small.}

Next, we consider the rest of the terms in \eqn{stresstensor10}.  Similarly to what we did in \eqn{newd2phi} to write $\partial^2 \Phi$ in terms of $\delta$, we can also write the other fields $\partial^2 \Psi$ and $\partial^2 \chi$ in terms of powers of $\delta$ (at least perturbatively, see  \cite{CLV1}).  Then, we can use the continuity equation \eqref{conteq1} to replace $\partial_j v^j$ with $\dot \delta$.  Because we are doing a one-loop computation, we only need to evaluate \eqn{stresstensor10} on the first-order fields, which means that the integrand on the right-hand side of \eqn{stresstensor10}  (without the stochastic piece) reduces to
\begin{align}
&   \kappa^{(\delta)}( a , a') \partial^i \delta ^{(1)} ( \xvec , a ') +  \kappa^{(v)} ( a , a') \partial^i  \delta ^{(1)} {}' ( \xvec , a ')   \ ,
\end{align}
where $\kappa^{(\delta)}$ is related to $\kappa^{(\Phi)}$, $\kappa^{(\Psi)}$, $\kappa^{(\chi)}$, and the $\mu$ parameters in \eqn{newd2phi} after making the replacements for $\partial^2 \Phi$, $\partial^2 \Psi$, and $\partial^2 \chi$ described above.  Again, the specific form of this relation is not important for what we describe here.  

Now, everything follows exactly as it does in other one-loop treatments \cite{Carrasco:2013mua, Lewandowski:2014rca, Lewandowski:2016yce}, and we can use the linear solution $\deltam^{(1)} ( \xvec , a ) = D_+(a)\, \deltam^{(1)} ( \xvec , a_i ) / D_+(a_i) $ to write 
\begin{align}
\begin{split}
 -  \left( \frac{1}{\rho_{\rm m}} \partial_j \tau^{ij} \right)_s^{(1)} (  \xvec , a )  &= \left( \int d a ' \left[  \kappa^{(\delta)}( a , a')  +  a^{-1} f_+ ( a )\kappa^{(v)} ( a , a')   \right] \frac{D_+(a')}{D_+(a)} \right) \partial^i \delta ^{(1)} ( \xvec , a) \\
 & = \kappa(a)\,  \partial^i \delta ^{(1)}  ( \xvec , a) \ ,
 \end{split}
\end{align}
where we have symbolically performed the integral over $a'$ to be left with a function of $a$ only.  By taking the Fourier transform of $\cH^{-2} \partial_i \left( \rho_{\rm m}^{-1}\partial_j \tau^{ij} \right)$, as it appears on the right-hand side of \eqn{eulereq1},  the first-order counterterm contribution to the right-hand side of \eqn{eulereq1} is
\be
- 9 \, ( 2 \pi ) c_{\delta ,1}^2 ( a ) \frac{k^2}{\knl^2} \deltam^{(1)} ( \kvec , a ) \;,
\ee
where we have defined $c_{\delta ,1}^2 (a ) \equiv - \kappa(a ) \knl^2 / ( 9 (2 \pi)^2 \cH(a)^2)$.\footnote{The factor of $9$ introduced here is simply a convention and is explained in Footnote~\ref{footnotelab}.}  This acts as a source term to the Euler equation in the same way that the nonlinear vertices do, so to find the counterterm contribution $\deltam^{(\rm ct)}$ to the perturbative expansion, we use the Green's functions from Sec.~\ref{greenfunsec}.  Putting this all together, we finally have\footnote{\label{footnotelab}We have introduced the relevant counterterm parameter that enters the power spectrum at one loop as
\be \label{defofc1}
 c_\delta ^2 ( a ) = \int^a d a' \, G^{\delta}_{2} ( a  , a' )  \, 9 \, c_{\delta ,1}^2 ( a' ) \frac{D_+ ( a ' )}{D_+(a)} \  . 
\ee
The factor of $9$ has been included because, approximating the quantities in \eqn{defofc1} with their Einstein de Sitter expressions and taking $c_{\delta ,1}^2 ( a ) \propto a^4$ as an indication (i.e. the time dependence in a scaling universe with $n=-2$ \cite{Pajer:2013jj}), we have that $  c_\delta ^2 ( a_0 ) \simeq c_{\delta ,1}^2 ( a_0)$.  
} 
\be
\deltam^{\rm ct} ( \kvec , a )  = - ( 2 \pi ) \,   c_\delta ^2 ( a ) \frac{k^2}{\knl^2} \frac{D_+(a)}{D_+(a_i)}  \deltam^{(1)} ( \kvec , a_i ) \ . 
\ee
That is, the counterterm contribution at one loop has the same functional form (in $k$) as the pure dark-matter case.  The counterterm $ c_\delta ^2 ( a )$, however, is expected to have a different value than in the pure dark matter case because the UV physics has been changed by the dark energy.  This gives a contribution to the power spectrum 
\be 
P^{\rm ct}_{13} ( k , a ) = - 2 ( 2 \pi )  c_\delta ^2 ( a ) \frac{k^2}{\knl^2} P_{11} ( k , a ) \ ,
\label{PS13cs}
\ee
which completes our calculation. Note that the nonlinear scale $\knl$ appearing above will in general be different than the corresponding scale in $\Lambda$CDM.\footnote{For instance, the modifications of the linear equations through $\mu_{\Phi} \neq 1$ affect the nonlinear scale. 
Using the expression for the linear power spectrum in a scaling universe \cite{Pajer:2013jj, Carrasco:2013sva, Carrasco:2013mua},
$P_{11} ( k , a  ) =  ({ D_+ (a)}/{D_+(a_0) )^2} {( 2 \pi /\knl )^3} ( { k }/{\knl} )^n $,
where $n \simeq -2$ near the nonlinear scale in the real universe,
the change in the nonlinear scale can be estimated by 
\be 
\label{knld}
\knl \sim \knl^{(\Lambda\text{CDM})}  \left( \frac{D_+^{(\Lambda\text{CDM})} ( a_i )}{ D_+^{(\Lambda\text{CDM})} ( a_0) }  \frac{ D_+ ( a_0 ) }{ D_+(a_i) } \right)^{- \frac{2}{n+3}} \ , 
\ee
where we have assumed the same initial conditions for the two theories. 
For $\mu_{\Phi} \simeq 1$,  the effect can be treated perturbatively in the linear equation of motion \eqref{lineaeq} and we have
\be 
\label{knld}
\knl \sim \knl^{(\Lambda\text{CDM})} \left( 1 -  \frac{3}{n+3} \int_{a_i}^{a_0} d a \,  G^{\delta, \Lambda\text{CDM}}_2 ( a_0 , a ) \big( \mu_{\Phi} ( a) - 1 \big)  \omegam ( a) \frac{D_+^{(\Lambda\text{CDM})} ( a ) }{D_+^{(\Lambda\text{CDM})} (a_0)} \right) \ , 
\ee
where $G^{\delta, \Lambda\text{CDM}}_2$ is  the Green's functions in $\Lambda$CDM.
Thus, a $\mu_{\Phi} > 1$  causes more linear growth and the nonlinear scale $\knl$ is smaller than the corresponding one in $\Lambda$CDM (non-linearities appear earlier and on larger scales).}

%
%
%
\section{Results} \label{resultssec}

While a complete exploration of the effects of different parameter choices is left for future work, in this section we would like to give a rough idea of the size of the effects that we are computing.

To use the above formalism to compute the one-loop power spectrum, we must assume a parametrization for the time dependence of the Hubble rate $H(a)$ and of the functions  $\alpha_I (a)$ appearing in \eqn{EFTaction_masses}. This in turn gives a time dependence to the functions $\mu_{\Phi}$, $\mu_{\Phi,2}$ and $\mu_{\Phi,22}$  that determine the modification of the LSS equations.  For simplicity,  we have chosen, 
\be
\alpha_I ( a ) = \alpha_{I,0} \frac{1 - \omegam (a) }{1 - \Omega_{\rm m,0}}\ ,  \qquad  \frac{\dot H}{H^2} (a)= -\frac32 \Omega_{\rm m} (a)  \;, \qquad \Omega_{\rm m} (a)= \frac{ \Omega_{\rm m,0}  }{  \Omega_{\rm m, 0}  + (1 - \Omega_{\rm m, 0}) (a/a_0)^3  } \;, 
\ee
where $\omegamz$ is the current  matter fraction.  This parameterization is such that the Hubble rate $H(a)$ and the matter fraction $\omegam (a)$ are as in $\Lambda$CDM, and the $\alpha_I (a)$ vanish during matter domination at early times.  This justifies the use of standard initial conditions, which we set at $a_i = 0.02$ using CAMB, assuming cosmological parameters $\Omega_{\rm m, 0} = 0.281$, $h = 0.697$, $A_\zeta^2 = 2.37 \times 10^{-9}$ and $n_s = 0.971$.

We display a sample  of our calculations in Fig.~\ref{alphabplot}.
In particular, in this figure we plot the ratio of the one-loop power spectrum with modifications of gravity and the one in $\Lambda$CDM at redshift $z=0$.
We consider the effects of three different dark-energy couplings, $\alphaB$, $\alphaq$, and $\alphac$, while setting the other $\alpha_I$ to zero. We restrict $\alpha_{\text{B},0}$ to negative values so that the speed of sound {squared} of scalar fluctuations is positive  (see e.g.~discussion in App. D of Ref.~\cite{DAmico:2016ntq}). Moreover, we consider the following combinations of parameters. First, $\alpha_{\rm B,0}=-0.9$ and $\alpha_{\rm B,0}=-0.5$ (thick solid blue and red lines, respectively) with $\alphaq= \alphac =0 $.  For these two cases we also show the linear power spectrum (thin dashed-dotted blue and red lines, respectively).
Then we consider $\alpha_{\rm B,0}=-0.5$ and several different values of $\alphaq \neq 0$ (with $\alphac=0$) and $\alphac \neq 0$ (with $\alphaq=0$). In particular we show the case $\alpha_{\rm V2,0} = 0.2$ (thick dashed violet),  $\alpha_{\rm V2,0} = 0.7$ (thick dashed green), and $\alpha_{\rm V1,0} =  0.7$ (thin solid green).\footnote{As explained above, $\alpha_{\rm V3} $ does not contribute to the one-loop power spectrum because it enters the fluid-like equations with a specific momentum dependence that vanishes when used to compute the one-loop power spectrum.  This vertex  contributes, however, to the one-loop bispectrum and two-loop power spectrum.}

To show these curves we need to assume a value of the speed of sound in the LSS counterterm in eq.~\eqref{PS13cs}. For illustration purposes we  take the representative value $ c_{\delta }^2 ( a_0 ) = 0.53 \, \knl^2 / ( 2\,  \unitsk )^2$,  which is the one measured in $\Lambda$CDM simulations in Ref.~\cite{Foreman:2015lca}. 
In the case of modified gravity, it is reasonable to think that the speed of sound  will differ from this value by something of order of the dark-energy couplings. Therefore, for two cases in the plot, we also show a shaded band corresponding to $ c_{\delta }^2 ( a ) \left( 1 \pm  \alpha_{\rm B,0}  \right)$, delimitating the plausible true value of the sound speed.

Not surprisingly, we see that the effect of $\alphaB$ enters most strongly at linear level and only has a sizable nonlinear effect when accompanied by a large change in the linear power spectrum.    
For instance, in the $\alpha_{\text{B},0} = -0.9$ case  the linear effect is about 16\% and the nonlinear effect is only about 2\%.  
On the other hand, since $\alphaq$ and $\alphac$ do not enter in the linear solution they only affect the power spectrum at mildly nonlinear scales.  For example, with $\alpha_{\text{B},0} = -0.5$,  $\alpha_{\text{V2},0} = 0.7$ produces about a 5\% change at $k = 0.1 \, \unitsk$.  Additionally, we see that, at least at one-loop, the effect of varying $ c_\delta ^2 ( a_0 )$ is essentially degenerate with changing $\alpha_{\text{V1},0}$ or $\alpha_{\text{V2},0}$.  

The plot also shows the difference between the linear and the nonlinear theory.  For example, the linear power spectrum in the $\alpha_{\text{B},0} = -0.5$ case is shown along with many different nonlinear power spectra.  In all cases, the nonlinear theory deviates from the linear theory by a few percent before around $ k \approx 0.1 \, \unitsk$.  Again, we leave a more specific exploration of the different effects of the EFTofDE couplings, their time parametrization, and the EFTofLSS counterterms, to future work.  In particular, it would be interesting to compare these results to N-body simulations to see how the EFTofLSS counterterms depend on values of the EFTofDE parameters.

%
%

%
%

\section{Conclusions}

{In this paper  we have combined the Effective Field Theory of Dark Energy and the Effective Field Theory of Large-Scale Structure to study dark matter clustering in the mildly nonlinear regime, for general dark energy and modified gravity models. The gravitational sector is described by the EFT action developed in a companion paper \cite{CLV1}, in terms   of six operators parametrized by time dependent functions, three of which start beyond linear order.

To understand how these nonlinear couplings affect the clustering of dark matter, in Sec.~\ref{Effective fluid} we  derived the effective fluid equations for dark matter,  by smoothing  the continuity and Euler equations. The smoothed Euler equation  includes a Newtonian potential sourced by the nonlinearities in the gravitational sector and 
an effective stress tensor generated by short modes.  {Because of the presence of a new field, it was important to derive the general way in which the UV modes enter the Euler equation.}
We found that the counterterms in the effective fluid equations  enter the Euler equation in the standard way as in $\Lambda$CDM, i.e.~as the divergence of a stress tensor.  This implies contributions to the power spectrum as $P^{\rm ct}_{13} \sim k^2 P_{11}$ and $P_{\rm stoch.} \sim k^4$. 
This was to be expected, however, since our system only has one propagating field in the Newtonian limit, so that there are no relative velocity effects which can generate counterterms without derivatives in the Euler equation.  Additionally, for the same reasons, we expect that the bias expansion takes the same general form as in $\Lambda$CDM, where one expands in powers and derivatives of $\partial_i \partial_j \Phi$, along with stochastic terms, integrated along the fluid trajectory. } 

We then explicitly constructed the one-loop solution in Sec.~\ref{Perturbation theory}.  Although the linear equation of motion of $\delta$ is scale independent, we must solve for the higher order time dependence with the exact Green's functions of the linear equations.  This leads to a rather lengthy expression for the one-loop power spectrum, since many diagrams with different time dependences must be summed together (Sec.~\ref{oneloopsec}).  This situation presents a potential problem: spurious IR and UV divergences in the individual loop terms, which must ultimately cancel when all of the terms are summed together, may not in practice fully cancel if the  time dependent coefficients are not computed accurately enough.  To avoid this problem, in App.~\ref{irsafesec} we have given the expressions for the IR\&UV-safe integrands, which have these spurious divergences removed at the level of the integrand.  In particular, we find no new spurious IR divergences, which shows that our system still obeys the dark-matter consistency relations {\cite{Peloso:2013zw, Kehagias:2013yd, Creminelli:2013mca}. 

In Sec.~\ref{resultssec}, we have presented a sampling of our results for the one-loop power spectrum, including the operators proportional to $\alphaB$, $\alphaq$, and $\alphac$.  We leave a more detailed study of the various parameter combinations, and their time parameterizations, to future work.  As shown in Fig.~\ref{alphabplot}, only the operators proportional to $\alphaq$, and $\alphac$ are unconstrained at linear order and can appreciably contribute to the mildly nonlinear regime. However, the recent tight bound on the difference between the speeds of gravitational waves and  light \cite{Monitor:2017mdv} has severely constrained $\alphaT$, $\alphaq$, $\alphac$ and $\alphaset$ \cite{Creminelli:2017sry,Ezquiaga:2017ekz,Baker:2017hug}. For $\alphaT=\alphaq=\alphac=\alphaset=0$ one finds that (see App.~\ref{app:definitions} for the explicit expressions)
\begin{align}
\begin{split}
\mu_{\Phi}- 1 & =\frac{(\alphaB -\alphaM)^2}{\nu} \;, \qquad \mu_{\Phi,2}  = \frac{\alphaM - 2 \alphaB}{2} \left( \frac{  \mu_\Phi - 1 }{\nu} \right)^{3/2} \;, \\
  \mu_{\Phi,22} & = {\frac{(\alpha_{\rm M} - 2 \alpha_{\rm B})^2(\mu_\Phi - 1)^2 }{2 \nu^3} } \;, \qquad \mu_{\Phi,3}=0 \;,
\end{split}
\end{align}
which shows that it is not possible to enhance the nonlinear contributions keeping the linear one small.
Thus, assuming these constraints, the nonlinear effects  in Horndeski theories can only come from the parameters $\alphaB$ and $\alphaM$ and be associated to  deviations in the linear predictions.  Even if the leading nonlinear effects are ruled out, it is still important to know, in a precision comparison to data, the nonlinear effects (which must be present because the dark-energy field nonlinearly realizes time diffeomorphisms) from the parameters that enter at linear level.

This work can be extended in several directions. For instance, one can now start using the machinery developed here to compute other  observables, such as the bispectrum, redshift space distortions and, ultimately, halo statistics in redshift space.
It would also be very useful to compare our predictions of the power spectrum to numerical N-body simulations and understand how the LSS speed of sound depends on the dark-energy parameters.  An obvious extension of this work is to include the operators of more general theories  that are  compatible with the constraints on the graviton speed, such as a subset of the GLPV Lagrangian \cite{Gleyzes:2014dya,Gleyzes:2014qga}.
Finally, one can also go beyond the quasi-static approximation and include the effects of the dark-energy field's propagation.}

%
\section*{Acknowledgements}
The authors are pleased to thank L.~Alberte, A.~Barreira, E.~Bellini, B.~Bose, P.~Creminelli, J.~Gleyzes, K.~Koyama,   F.~Schmidt, H.~Winther and M.~Zumalac{\'a}rregui for many useful discussions related to this project.  M.L.~and F.V.~are also pleased to thank the workshop DARK MOD and its Paris-Saclay funding, the organizers and participants for interesting discussions.  M.L.~acknowledges financial support from the Enhanced Eurotalents fellowship, a Marie Sklodowska-Curie Actions Programme. F.V.~acknowledges financial support from ``Programme National de Cosmologie and Galaxies'' (PNCG) of CNRS/INSU, France and  the French Agence Nationale de la Recherche under Grant ANR-12-BS05-0002.  The work of G.C. is supported by the Swiss National Science Foundation.

%
%

\newpage
\appendix

%
%

%
%
%

\section{Definitions and previous results}
\label{app:definitions}

The dimensionless symmetric matrix $A_{ab}$  in eq.~\eqref{total_action_pert_c} has components
\be
A_{ab}=\left(
\begin{array}{ccc}
0&1&-\alphaB\\
1&-1-\alphaT&\alphaM-\alphaT\\
-\alphaB& \alphaM-\alphaT&-\mathcal{C}_2
\end{array}
\right)\,,
\ee
with
\be
{\cal C}_2 = - \nu - \alphaB (\xi + \alphaT - \alphaM) \;, \qquad \xi = \alphaB (1+\alphaT) + \alphaT - \alphaM \;,
\ee
where $\nu$ is a positive, because of stability, parameter given by \cite{CLV1}
\be 
\nu \equiv -  \bigg\{  (1+\alphaB) \bigg[ \alphaB (1+\alphaT) + \alphaT  - \alphaM + \frac{\dot H}{H^2}   \bigg] + \frac{ \dot \alpha_{\rm B} }{H}  
+  \frac{ \bar \rho_{\rm m}}{2 M^2 H^2 }  \bigg\}  \;.
\ee
The dimensionless time-dependent arrays $B_{abc}$  and  $C_{abcd}$ parametrize the coupling strength between fields in eq.~\eqref{total_action_pert_c}. Their non-vanishing  elements     are 
\be
\begin{split} \label{nlcoeffs}
B_{123} &=  B_{312} =B_{231}=B_{213}=B_{321}=B_{132}={\btwo}\,,\\ 
B_{133} & = B_{313}=B_{331}= {\bone} \,, \\
B_{233} & = B_{323}=B_{332} =  {\cal C}_3     \,, \qquad B_{333}  ={\mathcal{C}_4}  \,, \\
 C_{1333} &=  C_{3133}= C_{3313}=C_{3331} =- \bthree  \,,\qquad 
 C_{3333}  = \mathcal{C}_5  \,,
\end{split}
\ee
where we have introduced the following combinations,
\begin{align}
{\cal C}_3  & \equiv - \alphaT - \btwo (1-\alphaM) - \btwo \frac{\dot H}{H^2} + \frac{ \dot \alpha_{\rm V2}}{H}  \;, \\
{\mathcal{C}_4} &\equiv - 4 \alphaB+ 2 \alphaM - 3 \alphaT   -( \bone +   \btwo )(1-\alphaM) - 3 \btwo \frac{\dot H}{H^2}  +  \frac{ \dotbone +  \dotbtwo }{H} \;,  \label{C4}\\
\mathcal{C}_5 & \equiv  3\left(  \alphaT -  \bone+\btwo + \bthree \right) -  (3 \btwo + \bthree )\alphaM + (3 \btwo + \bthree) \frac{\dot H}{H^2} - \frac{3 \dotbtwo  + \dotbthree}{H} \;. 
\end{align}

The functions $\mu_{\Phi} ( a )$, $\mu_{\Phi,2} ( a )$, $\mu_{\Phi,22} ( a )$, and $\mu_{\Phi , 3 }(a)$ in eq.~\eqref{sol_NL1}  are explicitely given, in terms of  the coefficients of the action \eqn{total_actionEFT}, by \cite{CLV1}
 \begin{align}\label{muphiexp}
 \begin{split}
 \mu_{\Phi} & = 1 + \alphaT + \frac{\xi^2}{\nu} \ , \hspace{.1in} \mu_\Psi = 1 + \frac{\xi \alphaB}{\nu} \  , \hspace{.1in} \mu_\chi = \frac{\xi}{\nu} \ , \\
\mu_{\Phi,2} & = { \frac{\mu_{\chi }}{4} }   \Big( 6  \mu_{\Phi} \mu_{\Psi} \btwo + 3 \mu_\chi \mu_\Phi  \bone  + 3 \mu_\chi \mu_\Psi   {\cal C}_3  + \mu_\chi^2 {\cal C}_4\Big) \;, \\
{ \mu_{\Phi,22} } & =  \frac{1}{8}   \bigg\{ 5 \mu_\Phi \mu_\chi^2 (\mu_\chi \bone  + 2 \mu_\Psi \btwo )^2 + 2 \mu_\chi^3 (3 \mu_\Psi {\cal C}_3 + \mu_\chi {\cal C}_4) (\mu_\chi \bone + 2 \mu_\Psi \btwo) \\
& \quad+\frac{1}{\nu} \big[ 2 \btwo \muphi ( 2 \mupsi- 1) +     2 \bone \muchi \muphi  + (3 \mupsi-1)  {\cal C}_3 \muchi  +  {\cal C}_4     \muchi^2 \big]^2    \bigg\}    \;,       \\
\mu_{\Phi,3} & =  { \frac{\mu_\chi^3}{12} }    \big( - 4 \mu_\Phi \bthree + \mu_\chi {\cal C}_5 \big) \;.
\end{split}
\end{align}

%
%
%

\section{Stress tensor} \label{stresstensorsec} 
In this appendix, we provide some explicit expressions for the higher order gravitational stress tensor that are relevant for the derivation of the Euler equation in Sec.~\ref{defluid}.  We start with the expressions for the $(0i)$ components.  In the quasi-static limit, the $(0i)$ part of the stress tensor can be written as a total derivative: $T^{\rm  (g)}{}^{0}{}_{i} = \partial_j t^j{}_{i}$.  The linear piece is
\be
\begin{split}
t^{\rm L}{}^{j}{}_i = \ & \delta{}^{j}{}_i M^2  \bigg\{  2 \dot \Psi + 2 (1+\alphaB) H \Phi  - 2 \alphaB H \dot \pi  + 2 \left(\dot H + \frac{\bar  \rho_{\rm m}}{2M^2} \right) \pi 
\bigg\} \; .
\end{split}
\ee
The second order piece is  
\be
\begin{split}
 t^{(2)}{}^{j}{}_i \equiv &  \frac{M^2}{a^2} \bigg[   H  \bigg(- 2\alphaB + \alphaM - \alphaT + \btwo \frac{\dot H}{H^2}   \bigg) \T_{2,ij} [ \pi,\pi] -  \alphaT \SSS_{2,ij} [\Psi,\pi] \\
& + (\bone - \btwo) \SSS_{2,ij} [\pi, \dot \pi - \Phi] + \frac{\btwo}{H} \big( \SSS_{2,ij} [\Psi, \dot \pi - \Phi] + \SSS_{2,ij} [ \pi,\dot \Psi + H \dot \pi]  \big)\bigg] 
\end{split} 
\ee
with
\be
 \T_{2,ij} [\varphi_a,\varphi_b] \equiv \partial_{(i} \varphi_a \partial_{j)} \varphi_b - \delta_{ij} \partial_k \varphi_a \partial_k \varphi_b \;, \qquad  \SSS_{2,ij} [ \varphi_a,\varphi_b] \equiv
\partial_i \varphi_a \partial_j \varphi_b - \delta_{ij} \partial_k \varphi_a \partial_k \varphi_b \; , 
\ee
{where we use the symmetrization normalization $V_{(ab)} = \half (V_{ab} + V_{ba})$}.  The expression for $t{}^{j}{}_i$ at cubic order is
\be
\begin{split}
t^{(3)}{}^{j}{}_i  =& -  \frac{M^2}{2\, a^4} \left( \bone -\btwo    -  \alphaM \btwo  -  \alphaT + \frac{\btwo \dot H}{ H^2} - \frac{ \dotbtwo }{H} \right) \left( \,  \mathcal{U}_{3,ij} [ \pi , \pi , \pi] + \frac{3}{2} \, \mathcal{V}_{3,ij} [ \pi , \pi , \pi] \right) \\
& + \frac{M^2}{4\, a^4 H} \left( -2 \btwo + \bthree \right) \big( 2 \, \mathcal{U}_{3,ij} [ \pi , \pi , \dot \pi - \Phi] - 2 \, \mathcal{U}_{3,ij} [ \pi ,  \dot \pi - \Phi, \pi] +2 \, \mathcal{U}_{3,ij} [  \dot \pi - \Phi, \pi , \pi ]  \\
  & \hspace{1.4in}   + \mathcal{V}_{3,ij} [ \pi, \pi, \dot \pi - \Phi ]  + \mathcal{V}_{3,ij} [ \pi, \dot \pi - \Phi, \pi ]  + \mathcal{V}_{3,ij} [  \dot \pi - \Phi , \pi , \pi ]  \big)
\end{split}
\ee
where 
\begin{align}
\mathcal{U}_{3,ij}[\varphi_a , \varphi_b , \varphi_c] & = \partial_j \varphi_a \partial_i \varphi_b \partial^2 \varphi_c - \partial_j \varphi_a \partial_k \varphi_b \partial_i \partial_k \varphi_c \ , \\
 \mathcal{V}_{3,ij}  [\varphi_a , \varphi_b , \varphi_c]  &= \partial_k \varphi_a \partial_k \varphi_b \partial_i \partial_j \varphi_c - \delta_{ij} \partial_k \varphi_a \partial_k \varphi_b \partial^2 \varphi_c \ . 
\end{align}

Next, we look at the $(ij)$ components.  At linear order, we have 
\be
\begin{split}
T^{\rm (g)}_{\rm L}{}^{j}{}_{i} =&  - M^2 \bigg\{  \bigg[ 2 \ddot \Psi + 2 (3+\alphaM) H \dot \Psi + 2 (1+\alphaB) H \dot \Phi  \\
& + \left( 2(1+\alphaB) \dot H + 2 \dot \alpha_{\rm B} H - \frac{\bar  \rho_{\rm m}}{M^2} +2(1+\alphaB)(3+\alphaM) H^2 \right) \Phi   - 2 (\alphaB H \dot \pi)^{\hbox{$\cdot$}} \\
&+ \left( 2 \dot H +\frac{\bar  \rho_{\rm m}}{M^2}  - 2 (3+\alphaM) \alphaB H^2 \right)  \dot \pi   + \left(2 \ddot H +2 (3 +\alphaM) H \dot H  \right) \pi  \bigg]  \delta{}^{j}{}_i \\
&  - \frac{1}{a^2}    (\partial_i \partial_j  - \delta{}^{j}{}_i \partial^2 ) \big[ \Phi +(\alphaM -\alphaT) \chi - (1+\alphaT) \Psi \big] \bigg\} \;.
\end{split}
\ee
At leading order in $\epsilon$, the nonlinear piece is 
\begin{align}
\label{TNLji}
T^{\rm (g)}_{\rm NL}{}^{j}{}_{i} =  \ &   \frac{M^2}{  2 H^2 a^4} \partial_k \bigg\{  -{\cal C}_3   \big[ 2   \partial_k \chi \partial_i \partial_j \chi  + \delta{}^{j}{}_i \partial_k (\partial \chi)^2 -  \delta{}^{j}{}_i   \partial_l  (\partial_k \chi \partial_l \chi ) - \delta_{(i}^k \partial_{j)}  (\partial \chi)^2 \big]  \\
& + 2 \btwo    \big[   \partial_k\chi \partial_{i} \partial_{j} \Phi  +  \partial_{ k} \Phi \partial_{i} \partial_{j} \chi  + \delta{}^{j}{}_i \partial_k ( \partial_l \chi \partial_l \Phi ) -  \delta{}^{j}{}_i  \partial_l(\partial_k \chi \partial_l \Phi ) -  \delta^k_{(i} \partial_{j)} ( \partial_l \chi \partial_l \Phi)  \big] \nonumber
\bigg\} \;.
\end{align}
We see that \eqn{TNLji} is ${\cal O}(\epsilon^0)$ (and, in fact, there are no terms of $\mathcal{O}(\epsilon^0)$ with three powers of the fields in $T^{\rm (g)}_{\rm NL}{}^{j}{}_{i}$).  This would be a dominant term in \eqn{deeulereq1}, but one can check that the divergence of the right-hand side vanishes identically, so that there are no contributions to the Euler equation at this order of spatial derivatives, as expected.

%
%
%
%

\section{Toolkit for the  one-loop calculation} \label{Apponeloop}

\subsection{Green's functions} \label{gfsec}

In this appendix, we provide the explicit formulae for the Green's functions used in Sec.~\ref{Perturbation theory} and throughout this work.  We use a slightly different notation than in \cite{Lewandowski:2016yce}, due to our different definition of $\Theta$. 

Using the perturbative expansion \eqref{dtGreen} and \eqref{dtGreen2} in the continuity and Euler equations \eqref{conteq1} and \eqref{eulereq1}, we find that the four Green's functions are specified by the following equations \cite{Lewandowski:2016yce}
 \begin{align}
 &a \frac{d G^{\delta}_{\sigma}(a,\ta)}{da}   -  G^{\Theta}_{\sigma}(a,\ta)=\lambda_{\sigma}\delta(a-\ta) \ ,  \label{Green} \\
 &a \frac{d G^{\Theta}_{\sigma}(a,\ta)}{da}    + \left( 1 + \frac{ a \cH'(a)}{\cH(a)} \right)  G^{\Theta}_{\sigma}(a,\ta) -  \mu_{\Phi } ( a ) \frac{3 \, \omegam (a )}{2   }  G^{\delta}_{\sigma}(a,\ta)    =(1-\lambda_{\sigma})\delta(a-\ta) \ ,
 \end{align}
where $\lambda_\sigma$ is $$\lambda_1=1 \quad \textmd{and} \quad \lambda_2=0,$$ $\sigma=1,2$, and $\delta(a-\ta)$ is the Dirac delta function. The retarded Green's functions satisfy the boundary conditions 
\be
\begin{split}  
& G^{\delta}_\sigma(a,\tilde a)  =  0 \quad \quad  \text{and}   \quad\quad G^{\Theta}_\sigma(a, \tilde a)=0  \quad \quad \text{for} \quad \quad \tilde a > a  \ , \\
 &G^\delta_\sigma ( \tilde a , \tilde a ) = \frac{\lambda_\sigma}{\tilde a}  \quad \hspace{.06in} \text{and} \hspace{.2in}  \quad G^{\Theta}_{\sigma} ( \tilde a  , \tilde a ) = \frac{(1 - \lambda_\sigma)}{\tilde a}  . \label{bound2}
\end{split}
\ee

We can then construct the Green's functions in the usual way using the linear solutions and the Heaviside step function, $\Theta_{\rm H} (a-\tilde a)$, and imposing the boundary conditions \eqref{bound2}.  This gives 
\begin{align}
&G^{\delta}_1(a,\ta) = \frac{1}{\ta W(\ta)}\bigg(\frac{d D_{-}(\ta)}{d\ta}  D_{+}(a)-\frac{d D_{+}(\ta)}{d\ta}D_{-}(a)\bigg)\Theta_{\rm H}(a-\ta)  \label{gdelta} \ , \\
&G^{\delta}_2(a,\ta)=  -  \frac{1}{ \ta^2 W(\ta)}    \bigg(  D_{-}(\ta)D_{+}(a)  - D_{+}(\ta)D_{-}(a)    \bigg)\Theta_{\rm H}(a-\ta)  \label{gdelta2} \ , \\
&G^{\Theta}_1(a,\ta)=   \frac{1}{ \ta W(\ta)}   \bigg(\frac{d D_{-}(\ta)}{d\ta}   \frac{a \, d D_{+}(a)}{d a}-\frac{d D_{+}(\ta)}{d\ta}     \frac{   a \, d D_{-}(a)}{d a}\bigg)\Theta_{\rm H}(a-\ta) \ ,\\
&G^{\Theta}_2(a,\ta)  =  - \frac{1}{ \ta^2 W(\ta)}    \bigg(  D_{-}(\ta)\frac{a \, d D_{+}(a)}{d a}  -  D_{+}(\ta)\frac{a\, d D_{-}(a)}{d a}     \bigg)\Theta_{\rm H}(a-\ta) \ ,   \label{gtheta}
\end{align}
where $W(\ta)$ is the Wronskian of $D_+$ and $D_-$ 
\be
W(\ta)=\frac{dD_{-}(\ta)}{d\ta}D_{+}(\ta)-\frac{d D_{+}(\ta)}{d\ta}D_{-}(\ta) \ .
\ee

For giving explicit formulae, it is useful  to define with a bar the part of the Green's functions \eqref{gdelta}---\eqref{gtheta} that do not contain the Heaviside function, i.e., 
\be
G^{\delta, \Theta}_{1,2}( a_1 , a_2 ) \equiv \bar G^{\delta,\Theta}_{1,2}( a_1 , a_2 ) \, \Theta_{\rm H}( a_1 - a_2) \ . \label{Gbar}
\ee 
For reference, the above Green's functions during matter domination, when $D_+ = a$ and $D_-~\propto~H \propto a^{-3/2}$, reduce to 
\begin{equation}
\begin{split}
&\bar G^{\delta}_1(a,\ta) = \frac{3 \, a }{5 \, \ta^2} + \frac{2 \, \ta^{1/2} }{5 \, a^{3/2}} 
\ , \hspace{.5in}  \bar G^{\delta}_2(a,\ta)=  \frac{2 \, a }{5 \, \ta^2} - \frac{2 \, \ta^{1/2} }{5 \, a^{3/2}}   \ , \\
&\bar G^{\Theta}_1(a,\ta)=   \frac{3 \, a }{5 \, \ta^2} - \frac{3 \, \ta^{1/2} }{5 \, a^{3/2}}  \ , \hspace{.5in} \bar G^{\Theta}_2(a,\ta)  =   \frac{2 \, a }{5 \, \ta^2} + \frac{3 \, \ta^{1/2} }{5 \, a^{3/2}}   \ .   \label{gthetamat}
\end{split}
\end{equation}

%
%
%

\subsection{Expressions for $\delta^{(2)}$ and $\delta^{(3)}$ }\label{delta23}
To make the notation more compact, because the $\mu$ parameters always appear with specific powers of $\omegam$ in the LSS equations, we define
\be \label{barmudef}
\hat \mu_{\Phi , 2 }   \equiv \mu_{\Phi , 2}   \left( \frac{ 3 \, \omegam}{2  } \right)^2  \ , \hspace{.1in}
\hat \mu_{\Phi , 22 }   \equiv \mu_{\Phi , 22}   \left( \frac{ 3 \, \omegam }{2  } \right)^3  \ , \hspace{.1in}
\hat \mu_{\Phi , 3 }   \equiv \mu_{\Phi , 3}   \left( \frac{ 3 \, \omegam  }{2  } \right)^3  \ . 
\ee

For computing the bispectrum or higher order power spectrum, it is useful to know the field contributions explicitly.  After defining the shorthand $\delta^{\rm in}_{\kvec } \equiv \delta^{(1)}( \kvec , a_i ) $, these are 
\be \label{delta2defapp}
\delta^{(2)} ( \kvec , a ) = \int_{\kvec_1} \int_{\kvec_2} \int_0^a d a_1 \sum_{i=1}^3  g^{(2)}_i ( a , a_1) y^{(2)}_i ( \kvec, \kvec_1 , \kvec_2) \delta^{\rm in}_{\kvec_1} \delta^{\rm in}_{\kvec_2} \ , 
\ee
where
\begin{align}
g^{(2)}_1 ( a , a_1) & \equiv \left( \frac{D_+(a_1)}{D_+(a_i)} \right)^2 \, f_+ ( a_1 ) \, G^\delta_1 ( a , a_1 )  \;, \\ 
g^{(2)}_2 ( a , a_1) & \equiv \left( \frac{D_+(a_1)}{D_+(a_i)} \right)^2 \, f_+ ( a_1 )^2  \, G^\delta_2 ( a , a_1 ) \;, \\ 
g^{(2)}_3 ( a , a_1) & \equiv \left( \frac{D_+(a_1)}{D_+(a_i)} \right)^2 \,  \hat \mu_{\Phi,2} ( a_1) \, G^\delta_2 ( a , a_1 ) \;, \label{g23} 
\end{align}
and
\begin{align}
y^{(2)}_1 ( \kvec, \kvec_1 , \kvec_2)  & \equiv ( 2 \pi)^3 \delta_D ( \kvec - \kvec_1 - \kvec_2 ) \alpha( \kvec_2 , \kvec_1) \;, \\
y^{(2)}_2 ( \kvec, \kvec_1 , \kvec_2)  & \equiv ( 2 \pi)^3 \delta_D ( \kvec - \kvec_1 - \kvec_2 ) \beta( \kvec_1 , \kvec_2) \;, \\
y^{(2)}_3 ( \kvec, \kvec_1 , \kvec_2)  & \equiv ( 2 \pi)^3 \delta_D ( \kvec - \kvec_1 - \kvec_2 ) \gamma_2 ( \kvec_1 , \kvec_2) \ . 
\end{align}

Similarly, for $\delta^{(3)}$ we have 
\be
\begin{split}
\delta^{(3)} ( \kvec , a )&  = \int_{\kvec_1} \int_{\kvec_2} \int_{\kvec_3} \int_0^a d a_1 \int_0^a d a_2 \sum_{i=1}^{12}  g^{(3)}_i ( a , a_1,a_2) y^{(3)}_i ( \kvec, \kvec_1 , \kvec_2, \kvec_3) \delta^{\rm in}_{\kvec_1} \delta^{\rm in}_{\kvec_2} \delta^{\rm in}_{\kvec_3} \\
& + \int_{\kvec_1} \int_{\kvec_2} \int_{\kvec_3}  \int_0^a d a_2 \sum_{i=13}^{14}  g^{(3)}_i ( a , a_2) y^{(3)}_i ( \kvec, \kvec_1 , \kvec_2, \kvec_3) \delta^{\rm in}_{\kvec_1} \delta^{\rm in}_{\kvec_2} \delta^{\rm in}_{\kvec_3}  \ . 
\end{split}
\ee
The terms with $i=13$ and $i=14$ only have one time integral because they come from the cubic vertex, which only needs one insertion of the Green's function to contribute to the cubic perturbation $\delta^{(3)}$.  Here, the dark-matter-only time-dependent coefficients are given by 
\be
\begin{split}
g^{(3)}_1 ( a , a_1,a_2) & \equiv  \frac{a_1 a_2 D_+(a_1) D_+'(a_1) D_+'(a_2)}{D_+(a_i)^3} G^\delta_1 ( a , a_2) G^\delta_1 ( a_2 , a_1 ) \ , \\
g^{(3)}_2 ( a , a_1,a_2) & \equiv  \frac{a_1^2 a_2 D_+'(a_1)^2 D_+'(a_2)}{D_+(a_i)^3} G^\delta_1 ( a , a_2) G^\delta_2 ( a_2 , a_1 ) \ , \\
g^{(3)}_3 ( a , a_1,a_2) & \equiv  \frac{a_1 D_+(a_1) D_+(a_2) D_+'(a_1) }{D_+(a_i)^3} G^\delta_1 ( a , a_2) G^\Theta_1 ( a_2 , a_1 ) \ , \\
g^{(3)}_4 ( a , a_1,a_2) & \equiv  \frac{a_1^2  D_+(a_2) D_+'(a_1)^2 }{D_+(a_i)^3} G^\delta_1 ( a , a_2) G^\Theta_2 ( a_2 , a_1 ) \ , \\
g^{(3)}_5 ( a , a_1,a_2) & \equiv  \frac{ a_1 a_2  D_+(a_1) D_+'(a_1) D_+'(a_2) }{D_+(a_i)^3} G^\delta_2 ( a , a_2) G^\Theta_1 ( a_2 , a_1 ) \ , \\
g^{(3)}_6 ( a , a_1,a_2) & \equiv  \frac{ a_1^2  a_2  D_+'(a_1)^2 D_+'(a_2) }{D_+(a_i)^3} G^\delta_2 ( a , a_2) G^\Theta_2 ( a_2 , a_1 )  \ ,
\end{split}
\ee
and the new coefficients are
\be
\begin{split}
g^{(3)}_7 ( a , a_1,a_2) & \equiv \hat \mu_{\Phi , 2} ( a_1 )  \frac{ D_+(a_1)^2 D_+(a_2) }{D_+(a_i)^3} G^\delta_1 ( a , a_2) G^\Theta_2 ( a_2 , a_1 ) \ , \\
g^{(3)}_8 ( a , a_1,a_2) & = \hat \mu_{\Phi , 2} ( a_2 )  \frac{ 2 a_1 D_+(a_1)  D_+(a_2) D_+'(a_1) }{D_+(a_i)^3} G^\delta_2 ( a , a_2) G^\delta_1 ( a_2 , a_1 ) \ , \\
g^{(3)}_9 ( a , a_1,a_2) & \equiv \hat \mu_{\Phi , 2} ( a_2 )  \frac{ 2 a_1^2 D_+(a_2)  D_+'(a_1)^2 }{D_+(a_i)^3} G^\delta_2 ( a , a_2) G^\delta_2 ( a_2 , a_1 ) \ , \\
g^{(3)}_{10} ( a , a_1,a_2) & \equiv \hat \mu_{\Phi , 2} ( a_1 )  \frac{  a_2 D_+(a_1)^2  D_+'(a_2) }{D_+(a_i)^3} G^\delta_1 ( a , a_2) G^\delta_2 ( a_2 , a_1 ) \ , \\
g^{(3)}_{11} ( a , a_1,a_2) & \equiv \hat \mu_{\Phi , 2} ( a_1 )  \frac{  2 a_2 D_+(a_1)^2  D_+'(a_2) }{D_+(a_i)^3} G^\delta_2 ( a , a_2) G^\Theta_2 ( a_2 , a_1 ) \ , \\
g^{(3)}_{12} ( a , a_1,a_2) & \equiv \hat \mu_{\Phi , 2} ( a_1 ) \hat \mu_{\Phi , 2} ( a_2 )  \frac{  2  D_+(a_1)^2  D_+(a_2) }{D_+(a_i)^3} G^\delta_2 ( a , a_2) G^\delta_2 ( a_2 , a_1 ) \ , \\
g^{(3)}_{13} ( a , a_2) & \equiv \hat \mu_{\Phi , 22} ( a_2 )  \frac{   D_+(a_2)^3 }{D_+(a_i)^3} G^\delta_2 ( a , a_2)  \ , \\
g^{(3)}_{14} ( a , a_2) & \equiv \hat \mu_{\Phi , 3} ( a_2 )  \frac{   D_+(a_2)^3 }{D_+(a_i)^3} G^\delta_2 ( a , a_2)  \ .
\end{split}
\ee

The dark-matter-only momentum dependent coefficients are given by 
\be
\begin{split}
 y^{(3)}_1 ( \kvec, \kvec_1 , \kvec_2, \kvec_3) & \equiv ( 2 \pi)^3 \delta_D ( \kvec - \kvec_1 - \kvec_2 - \kvec_3 ) \,\alpha ( \kvec_2 , \kvec_1 + \kvec_3 )\, \alpha ( \kvec_3 , \kvec_1 ) \ , \\
  y^{(3)}_2 ( \kvec, \kvec_1 , \kvec_2, \kvec_3) & \equiv ( 2 \pi)^3 \delta_D ( \kvec - \kvec_1 - \kvec_2 - \kvec_3 ) \,\alpha ( \kvec_2 , \kvec_1 + \kvec_3 ) \, \beta ( \kvec_1 , \kvec_3 ) \ , \\
   y^{(3)}_3 ( \kvec, \kvec_1 , \kvec_2, \kvec_3) & \equiv ( 2 \pi)^3 \delta_D ( \kvec - \kvec_1 - \kvec_2 - \kvec_3 )\, \alpha ( \kvec_2 + \kvec_3 , \kvec_1 )\, \alpha ( \kvec_3 , \kvec_2 ) \ , \\
y^{(3)}_4 ( \kvec, \kvec_1 , \kvec_2, \kvec_3) & \equiv ( 2 \pi)^3 \delta_D ( \kvec - \kvec_1 - \kvec_2 - \kvec_3 ) \,\alpha ( \kvec_2 + \kvec_3 , \kvec_1 ) \, \beta ( \kvec_2 , \kvec_3 ) \ , \\   
y^{(3)}_5 ( \kvec, \kvec_1 , \kvec_2, \kvec_3) & \equiv ( 2 \pi)^3 \delta_D ( \kvec - \kvec_1 - \kvec_2 - \kvec_3 ) \, 2 \,  \beta (  \kvec_1  + \kvec_3 ,  \kvec_2  ) \, \alpha ( \kvec_3 , \kvec_1 ) \ , \\   
y^{(3)}_6 ( \kvec, \kvec_1 , \kvec_2, \kvec_3) & \equiv ( 2 \pi)^3 \delta_D ( \kvec - \kvec_1 - \kvec_2 - \kvec_3 )\, 2 \,   \beta (  \kvec_1  + \kvec_3 ,  \kvec_2  ) \, \beta ( \kvec_1 , \kvec_3 ) \ ,
\end{split}
\ee
and the new coefficients are
\be
\begin{split}
y^{(3)}_7 ( \kvec, \kvec_1 , \kvec_2, \kvec_3) & \equiv ( 2 \pi)^3 \delta_D ( \kvec - \kvec_1 - \kvec_2 - \kvec_3 ) \, \alpha ( \kvec_2 + \kvec_3, \kvec_1  ) \, \gamma_2 ( \kvec_2 , \kvec_3 ) \ , \\
y^{(3)}_8 ( \kvec, \kvec_1 , \kvec_2, \kvec_3) & \equiv ( 2 \pi)^3 \delta_D ( \kvec - \kvec_1 - \kvec_2 - \kvec_3 ) \,\alpha (  \kvec_3, \kvec_1  ) \, \gamma_2 ( \kvec_1 + \kvec_3 , \kvec_2 ) \ , \\
y^{(3)}_9 ( \kvec, \kvec_1 , \kvec_2, \kvec_3) & \equiv ( 2 \pi)^3 \delta_D ( \kvec - \kvec_1 - \kvec_2 - \kvec_3 ) \,\beta (  \kvec_1, \kvec_3  ) \, \gamma_2 ( \kvec_1 + \kvec_3 , \kvec_2 ) \ , \\
y^{(3)}_{10} ( \kvec, \kvec_1 , \kvec_2, \kvec_3) & \equiv ( 2 \pi)^3 \delta_D ( \kvec - \kvec_1 - \kvec_2 - \kvec_3 ) \,\alpha (  \kvec_2, \kvec_1 + \kvec_3  ) \, \gamma_2 ( \kvec_1 , \kvec_3  ) \ , \\
y^{(3)}_{11} ( \kvec, \kvec_1 , \kvec_2, \kvec_3) & \equiv ( 2 \pi)^3 \delta_D ( \kvec - \kvec_1 - \kvec_2 - \kvec_3 )\, \beta (   \kvec_1 + \kvec_3 ,\kvec_2 ) \, \gamma_2 ( \kvec_1 , \kvec_3  ) \ , \\
y^{(3)}_{12} ( \kvec, \kvec_1 , \kvec_2, \kvec_3) & \equiv ( 2 \pi)^3 \delta_D ( \kvec - \kvec_1 - \kvec_2 - \kvec_3 ) \,\gamma_2 (   \kvec_1 + \kvec_3 ,\kvec_2 ) \, \gamma_2 ( \kvec_1 , \kvec_3  )  \ , \\
y^{(3)}_{13} ( \kvec, \kvec_1 , \kvec_2, \kvec_3) & \equiv ( 2 \pi)^3 \delta_D ( \kvec - \kvec_1 - \kvec_2 - \kvec_3 ) \,\gamma_2 (   \kvec_2, \kvec_1 + \kvec_3  ) \, \gamma_2 ( \kvec_1 , \kvec_3  ) \ , \\
y^{(3)}_{14} ( \kvec, \kvec_1 , \kvec_2, \kvec_3) & \equiv ( 2 \pi)^3 \delta_D ( \kvec - \kvec_1 - \kvec_2 - \kvec_3 ) \,\gamma_3 (   \kvec_2, \kvec_1 , \kvec_3  )  \ .
\end{split}
\ee

\subsection{Comparison with standard perturbation theory}

Using the explicit expressions for $\delta^{(2)}$ above, we can compare the new kernel with the well known kernel from dark-matter perturbation theory, which is given by  
\begin{align}
&\delta^{(2)} ( \kvec , a ) = \int_{\kvec_1} \int_{\kvec_2} ( 2 \pi)^3 \delta_D ( \kvec - \kvec_1 - \kvec_2  ) \,  \mathcal{F}_2 ( \kvec_1 , \kvec_2, a  ) \delta^{\rm in}_{\kvec_1} \delta^{\rm in}_{\kvec_2} \ .
\end{align}
In the Einstein de Sitter  limit the kernel ${\cal F}_2$ is given by \cite{Bernardeau:2001qr}
\be
\mathcal{F}^{\text{EdS}}_2 ( \kvec_1 , \kvec_2, a )  =   \frac{D_+(a)^2}{D_+(a_i)^2} \bigg[ \frac{5}{7} + \frac{ \hat k_1 \cdot \hat k_2 }{2} \left( \frac{k_1 }{k_2 } + \frac{ k_2 }{k_1} \right) + \frac{2}{7} \big( \hat k_1 \cdot \hat k_2 \big)^2   \bigg] \ . \label{EdS}
\ee
More generally, 
our formula \eqref{delta2defapp} for $\delta^{(2)}$ gives 
\begin{equation} \label{f2kernel}
\mathcal{F}_2 ( \kvec_1 , \kvec_2, a )  = \AAA  + \CCC  + \big( \hat k_1 \cdot \hat k_2 \big)   \frac{\AAA  + \BBB }{2} \left( \frac{k_1}{k_2} + \frac{k_2}{k_1} \right) \nonumber  + \big( \hat k_1 \cdot \hat k_2 \big)^2 \left(  \BBB   - \CCC  \right) \ , 
\end{equation}
where
\begin{equation} \label{somedefinitions}
\AAAA_i(a)  \equiv \int_0^a \, d a' \, g_i^{(2)} ( a , a' ) \;.
\end{equation}
By using the expressions above it can be shown that the coefficient of the monopole, $(\AAA +\BBB)/2$, is not altered by the presence of the dark energy and modified gravity and remains the same as in eq.~\eqref{EdS}.
On the other hand, the coefficients in front of $\big( \hat k_1 \cdot \hat k_2 \big)^0$ and $\big( \hat k_1 \cdot \hat k_2 \big)^2$  are altered explicitly by $\CCC$, 
which is the only term that depends on $\mu_{\Phi,2}$, see eq.~\eqref{g23},
and implicitly by $\mu_{\Phi  }$ in the expressions for $\AAA$ and $\BBB$.  A similar result holds for the monopole in the expression for the second-order velocity divergence $\theta^{(2)}$.

%
%
%

\subsection{Expressions for $P_{1\text{-loop}}$ } \label{expexp}

In this appendix, we provide the explicit formulae for the contributions to the one-loop power spectrum presented in Sec.~\ref{oneloopsec}.

We start with the $(22)$ type terms in eq.~\eqref{p22foryou}.
The dark-matter-only momentum-dependent functions are
\be
\begin{split} \label{longlist3}
F^{(22)}_1 ( \kvec , \qvec ) & = 2 \, \alpha_s( \kvec - \qvec , \qvec )^2  \, P^{\rm in}_{|\kvec - \qvec|}\, P^{\rm in}_{q} \ ,   \\
F^{(22)}_2 ( \kvec , \qvec ) & = 2 \, \alpha_s( \kvec - \qvec , \qvec ) \, \beta ( \kvec - \qvec , \qvec) \, P^{\rm in}_{|\kvec - \qvec|}\, P^{\rm in}_{q} \ ,\\
F^{(22)}_3 ( \kvec , \qvec )  & = 2 \, \alpha_s( \kvec - \qvec , \qvec ) \, \beta ( \kvec - \qvec , \qvec)  \,P^{\rm in}_{|\kvec - \qvec|}\, P^{\rm in}_{q} \ ,\\
F^{(22)}_4 ( \kvec , \qvec ) & = 2 \, \beta( \kvec - \qvec , \qvec ) \, \beta ( \kvec - \qvec , \qvec) \,P^{\rm in}_{|\kvec - \qvec|}\, P^{\rm in}_{q} \ ,  
\end{split}
\ee
and the new $(22)$ terms are
\be
\begin{split} \label{longlist32appp}
F^{(22)}_5 ( \kvec , \qvec ) & = 2\, \alpha_s ( \kvec - \qvec , \qvec ) \,  \gamma_2 ( \kvec - \qvec , \qvec)  \, P^{\rm in}_{|\kvec - \qvec|}\, P^{\rm in}_{q} \ ,   \\
F^{(22)}_6 ( \kvec , \qvec ) & = 2 \, \beta ( \kvec - \qvec , \qvec ) \,  \gamma_2 ( \kvec - \qvec , \qvec)  \, P^{\rm in}_{|\kvec - \qvec|}\, P^{\rm in}_{q} \ ,    \\
F^{(22)}_7 ( \kvec , \qvec ) & = 2 \, \gamma_2 ( \kvec - \qvec , \qvec ) \,  \gamma_2 ( \kvec - \qvec , \qvec)  \, P^{\rm in}_{|\kvec - \qvec|}\, P^{\rm in}_{q}    \ .  
\end{split}
\ee
In the above $\alpha_s ( \qvec_1 , \qvec_2) = \half ( \alpha( \qvec_1 , \qvec_2 ) + \alpha ( \qvec_2 , \qvec_1) )$.  To get the compact forms in  \eqn{longlist42}, we have used the properties that $\alpha_s$ and $\beta$ and $\gamma_2$ are symmetric and switched the variable of integration from $\qvec$ to $ - \qvec $ in some terms. 

The  time-dependent coefficients are given by 
\begin{align}
T^{(22)}_i ( a , a_1 , a_2) &= \tilde T^{(22)}_i ( a , a_1 , a_2) + \tilde T^{(22)}_i ( a , a_2 , a_1)  \ , 
\end{align}
where the coefficients from the dark-matter-only theory are given by 
\be
\begin{split}
\tilde T^{(22)}_1 ( a , a_1 , a_2) &= K( a_2 , a_1 , a_2)  \,   \bar G^\delta_1 ( a , a_1) \bar G^\delta_1 ( a , a_2) \ ,\\
\tilde T^{(22)}_2 ( a , a_1 , a_2) &= K( a_2 , a_1 , a_2)  \,  f_+(a_1)  \bar G^\delta_1 ( a , a_2)   \bar G^\delta_2 ( a , a_1) \ ,  \\
\tilde T^{(22)}_3 ( a , a_1 , a_2) &=  K( a_2 , a_1 , a_2)  \,  f_+(a_2)  \bar G^\delta_1 ( a , a_1)   \bar G^\delta_2 ( a , a_2) \ ,\\
\tilde T^{(22)}_4 ( a , a_1 , a_2) & = K( a_2 , a_1 , a_2)  \, f_+(a_1) \, f_+(a_2) \,   \bar G^\delta_2 ( a , a_1) \bar G^\delta_2 ( a , a_2) \ ,
\end{split}
\ee
and the new coefficients are given by 
\be
\begin{split}
\tilde T^{(22)}_5 ( a , a_1 , a_2) &= 2 \hat \mu_{\Phi , 2} ( a_1) K ( a_2 , a_1 , a_2 ) \, f_+(a_1)^{-1} \,       \bar G^\delta_2 ( a , a_1) \bar G^\delta_1 ( a , a_2) \; ,         \\
\tilde T^{(22)}_6 ( a , a_1 , a_2) &=  2 \hat \mu_{\Phi , 2} ( a_1)  K ( a_2 , a_1 , a_2 ) \, f_+(a_1)^{-1} \, f_+(a_2) \,     \bar G^\delta_2 ( a , a_1) \bar G^\delta_2 ( a , a_2)       \;,      \\
\tilde T^{(22)}_7 ( a , a_1 , a_2) &=          \hat \mu_{\Phi , 2} ( a_1)   \hat \mu_{\Phi , 2} ( a_2)     K ( a_2 , a_1 , a_2 ) f_+(a_1)^{-1} \, f_+(a_2)^{-1} \,      \bar G^\delta_2 ( a , a_1) \bar G^\delta_2 ( a , a_2)    \ ,
\end{split}
\ee
where the common factor $K$ is given by 
\be
K( a , a_1 , a_2 ) = \frac{ a_1 a_2 D_+(a ) D_+ ( a_1) D'_+(a_1) D'_+(a_2)}{D_+(a_i)^4} \ ,
\ee
and $f_\pm ( a ) \equiv a D_\pm ' ( a ) / D_\pm ( a ) $.

Now we move on to the $(13)$ type terms present in eq.~\eqref{p13foryou}.  
First, the dark-matter-only momentum functions are
\be
\begin{split} \label{longlist1}
F^{(13)}_1 ( \kvec , \qvec) & = 4 \,  \alpha_s ( \kvec , \qvec) \, \alpha ( - \qvec , \kvec + \qvec) \, P^{\rm in}_{k}\, P^{\rm in}_{q} \ , \\
F^{(13)}_2 ( \kvec , \qvec) & = 4 \,  \beta ( \kvec , \qvec) \, \alpha ( - \qvec , \kvec + \qvec) \, P^{\rm in}_{k}\, P^{\rm in}_{q}  \ , \\ 
F^{(13)}_3 ( \kvec , \qvec) & = 4 \,  \alpha_s ( \kvec , \qvec) \, \alpha ( \kvec + \qvec , - \qvec)\, P^{\rm in}_{k}\, P^{\rm in}_{q}   \ , \\
F^{(13)}_4 ( \kvec , \qvec) & = 4 \,  \beta ( \kvec , \qvec) \, \alpha (  \kvec + \qvec, -\qvec) \, P^{\rm in}_{k}\, P^{\rm in}_{q} \ , \\
F^{(13)}_5 ( \kvec , \qvec) & =  4 \times 2 \,  \alpha_s ( \kvec , \qvec) \, \beta ( - \qvec , \kvec + \qvec)\, P^{\rm in}_{k}\, P^{\rm in}_{q}  \ , \\
F^{(13)}_6 ( \kvec , \qvec) & = 4 \times 2 \,  \beta ( \kvec , \qvec) \, \beta ( - \qvec , \kvec + \qvec) \, P^{\rm in}_{k}\, P^{\rm in}_{q}  \ , 
\end{split}
\ee
and the new $(13)$ terms are
\be
\begin{split}  
F^{(13)}_7 ( \kvec , \qvec) & = 4 \,  \gamma_2 ( \kvec , \qvec) \, \alpha (  \kvec + \qvec , - \qvec ) \, P^{\rm in}_{k}\, P^{\rm in}_{q}  \ , \\
F^{(13)}_8 ( \kvec , \qvec) &= F^{(13)}_{10} ( \kvec , \qvec)  = F^{(13)}_{11} ( \kvec , \qvec)  = 4 \,  \alpha_s ( \kvec , \qvec) \, \gamma_2 (  \kvec + \qvec , - \qvec ) \, P^{\rm in}_{k}\, P^{\rm in}_{q}  \ , \\
F^{(13)}_{9} ( \kvec , \qvec) & = 4 \,  \gamma_2 ( \kvec , \qvec) \, \beta ( - \qvec , \kvec + \qvec ) \, P^{\rm in}_{k}\, P^{\rm in}_{q}  \ . \label{longlist42}
\end{split}
\ee
 We would like to point out that, although the contraction of $\delta^{(1)}$ with $\delta^{(3)}$ should naively produce fourteen terms, three of them are zero after the contraction.  In particular, the vertex that would be proportional to $\mu_{\Phi,3}$ does not contribute to the one-loop power spectrum because $\gamma_3 ( \kvec , \qvec,-\qvec) = 0$.  However, this vertex will contribute to the two-loop power spectrum, the one-loop bispectrum, and the tree level trispectrum.

Using eq.~\eqref{Gbar}, the dark-matter-only time dependent coefficients are
\be
\begin{split}
T^{(13)}_1 ( a , a_1 , a_2) & = K( a , a_1 , a_2 )\,  \bar G^\delta_1 ( a , a_2) \bar G^\delta_1 ( a_2 , a_1)  \ ,\\
T^{(13)}_2 ( a , a_1 , a_2) & = K( a , a_1 , a_2) \,  f_+(a_1) \, \bar G^\delta_1 ( a , a_2) \bar G^\delta_2 ( a_2 , a_1) \ , \\
T^{(13)}_3 ( a , a_1 , a_2) & = K( a , a_1 , a_2) \,f_+ ( a_2)^{-1} \,   \bar G^\delta_1 ( a , a_2) \bar G^\Theta_1 ( a_2 , a_1) \ ,\\
T^{(13)}_4 ( a , a_1 , a_2) & = K( a , a_1 , a_2) \,   f_+(a_1)  \,f_+ ( a_2)^{-1} \,  \bar G^\delta_1 ( a , a_2) \bar G^\Theta_2 ( a_2 , a_1) \ ,\\
T^{(13)}_5 ( a , a_1 , a_2) & = K( a , a_1 , a_2) \,  \bar G^\delta_2 ( a , a_2) \bar G^\Theta_1 ( a_2 , a_1)\ ,\\
T^{(13)}_6 ( a , a_1 , a_2) & = K( a , a_1 , a_2) \,  f_+(a_1) \, \bar G^\delta_2 ( a , a_2) \bar G^\Theta_2 ( a_2 , a_1) \ 
\end{split}
\ee
The new time dependent coefficients are given by 
\be
\begin{split}
T^{(13)}_7 ( a , a_1 , a_2) & = \hat \mu_{\Phi,2}( a_1)  K( a , a_1 , a_2 ) \, f_+(a_1)^{-1} \, f_+(a_2)^{-1}\,   \bar G^\delta_1 ( a , a_2) \bar G^\Theta_2 ( a_2 , a_1)  \;,  \\
T^{(13)}_8 ( a , a_1 , a_2) & = 2 \,  \hat \mu_{\Phi,2}( a_2) K( a , a_1 , a_2 )  \, f_+(a_2)^{-1}\,  \bar G^\delta_2 ( a , a_2) \bar G^\delta_1 ( a_2 , a_1)   \;,   \\
T^{(13)}_{9} ( a , a_1 , a_2) & =2 \,  \hat \mu_{\Phi,2}( a_1) K( a , a_1 , a_2 ) \, f_+(a_1)^{-1} \,  \bar G^\delta_2 ( a , a_2) \bar G^\Theta_2 ( a_2 , a_1) \;,   \\
T^{(13)}_{10} ( a , a_1 , a_2) & = 2 \, \hat \mu_{\Phi,2}( a_1) \hat \mu_{\Phi,2}( a_2)  K( a , a_1 , a_2 ) \, f_+(a_1)^{-1} \, f_+(a_2)^{-1} \,    \bar G^\delta_2 ( a , a_2) \bar G^\delta_2 ( a_2 , a_1)  \;, \\ 
\end{split}
\ee
and
\be
T^{(13)}_{11} ( a , a_2)  = \hat \mu_{\Phi,22}( a_2) \left( \frac{ D_+(a) D_+(a_2)^3}{D_+(a_i)^4} \right) \bar G^\delta_2 ( a , a_2)    \, .
\ee

%
%
%

 \subsection{Infrared- and ultraviolet-safe one-loop power spectrum} \label{irsafesec}

To derive the IR\&UV-safe power spectrum, it is helpful to separate the discussion in two parts. First, we will discuss the divergences  of the standard vertices  that are present in $\Lambda$CDM, i.e.~for $\mu_{\Phi,2}=\mu_{\Phi,22}=\mu_{\Phi,3}=0$. In this case the discussion is the same as the one of Ref.~\cite{Lewandowski:2017kes}, because the momentum dependent kernels are the same. 
Further below, we will address the non-standard terms that are coming from the nonlinear modify-gravity vertices, which instead are new. As we will see, these non-standard pieces can be straightforwardly treated as they do not give rise to  divergences that are not removed by the standard procedure. 

We remind that $P_{1\text{-loop}}= P_{22} + P_{13}$ and that eqs.~\eqref{p22foryou} and \eqref{p13foryou} express $P_{22}$ and $P_{13}$ in terms of their integrands $p_{22}(a, a_1 , a_2 ; \kvec , \qvec)$, $p_{13}^{(2)}(a, a_1 , a_2 ; \kvec , \qvec)$ and $p_{13}^{(1)}(a , a_2 ; \kvec , \qvec)$, which are defined in eq.~\eqref{p22text2} in terms of the kernels $F^{(22)}_i ( \kvec , \qvec )$ and $F^{(13)}_i ( \kvec , \qvec )$. 
The divergencies can be tracked in the way  the kernels $F^{(22)}_i ( \kvec , \qvec )$ and $F^{(13)}_i ( \kvec , \qvec )$ behave in the IR or UV limit.

Let us start by discussing the IR limit, i.e.~the limit $q / k \rightarrow 0$ and $\kvec \rightarrow \qvec$, of the standard contributions, where $\vec k$ is the power spectrum wavenumber, $\vec q$ is the one running in the loop and $k \equiv | \vec k|$ and $q \equiv | \vec q|$. As explained in Sec.~\ref{oneloopsec}  the standard contributions to $p_{22}$ and $p_{13}$ come, respectively, from  $i=1, \ldots,4$ and $i=1, \ldots,6$ in the sums in eq.~\eqref{p22text2}.  In the limit $q / k \rightarrow 0$ the kernels generically have the form 
\begin{align}
F^{(22)}_i ( \kvec , \qvec )  & = \left(  \frac{\mu^2}2 \frac{k^2}{q^2}  + \big(b_{i,1}^{\rm IR} \; \mu + b_{i,2}^{\rm IR} \; \mu^3 \big) \frac{k}{q} + \mathcal{O} \left( {k^0}/{q^0} \right) \right) P_{k}^{\rm in} P_{q}^{\rm in}  \qquad i=1,2,3,4\;, \\
F^{(13)}_i ( \kvec , \qvec )  & = \left(  - 2 {\mu^2} \frac{k^2}{q^2}   + \mathcal{O} \left( {k^0}/{q^0} \right) \right) P_{k}^{\rm in} P_{q}^{\rm in} \qquad i =1,2,5,6 \;, \\
F^{(13)}_i ( \kvec , \qvec )  & =  \mathcal{O} \left( {k^0}/{q^0} \right)  P_{k}^{\rm in} P_{q}^{\rm in} \qquad i =3,4 \;,
\end{align}
where $b_{i,1}^{\rm IR}$ and $b_{i,2}^{\rm IR}$ are numerical coefficients whose exact value is irrelevant here and $\mu \equiv \hat k \cdot \hat q$. The equivalence principle guarantees that both of the leading terms above, proportional to $k^2 / q^2$ and $k/q$, must cancel in the final expression for the equal-time power spectrum, i.e.~after adding together the contributions in \eqn{p22foryou} and \eqn{p13foryou}.

Notice  that $F^{(22)}_i ( \kvec , \qvec ) = F^{(22)}_i ( \kvec , \kvec - \qvec ) $, so that any IR divergence from $\qvec \rightarrow 0$ has  a corresponding IR divergence for $\qvec \rightarrow \kvec$.  Following \cite{Carrasco:2013sva}, using this property we can map the divergence at $\qvec \rightarrow \kvec$ to $\qvec \rightarrow 0 $ by writing the momentum loop integral as 
\be
 \int \momspmeas{q} \, \,  F^{(22)}_i ( \kvec , \qvec ) = 2  \int \momspmeas{q} \, \,   \Big( F^{(22)}_i ( \kvec , \qvec)\Theta_{\rm H} \big( | \kvec - \qvec | - q \big) 
 + F^{(22)}_i ( \kvec , -\qvec)\Theta_{\rm H} \big( | \kvec + \qvec | - q \big) \Big) \;, 
 \label{rewritingkq}
\ee
so that the integration does not involve the region $\qvec \approx \kvec$ any longer.  This means that we only have to consider the $q / k \rightarrow 0$ limit of $p_{22}$. This proceedure has also the advantage of cancelling  the subleading divergences in $k/q$, which are odd in $\mu$ (i.e.~in $\vec q$) and so they manifestly cancel in the integrand because of the antisymmetrization over $\qvec$.

 Thus, the IR that we need to subtract out are now  given by
\be
  \begin{split}  
   F^{(22)}_{i,{\rm IR}} ( \kvec , \qvec ) &=  \frac{ \mu^2 }{2}  \frac{k^2}{q^2}   P^{\rm in}_{\kvec} P^{\rm in}_{\qvec}  \qquad i=1,2,3,4\;, \\
   F^{(13)}_{i , \text{IR}} ( \kvec , \qvec) &=     - 2 \mu^2  \frac{k^2}{q^2} P^{\rm in}_{\kvec} P^{\rm in}_{\qvec} \qquad i =1,2,5,6 \;, \qquad
 F^{(13)}_{i , \text{IR}} ( \kvec , \qvec)  = 0 \qquad i =3,4 \, ,   \end{split}
 \label{IRcont}
 \ee
 where $F^{(22,13)}_{i , \text{IR}}( \kvec , \qvec )  = \lim_{q \to 0} F^{(22,13)}_{i }( \kvec , \qvec ) $.
It is straightforward to check (see \cite{Lewandowski:2017kes} for details) that the sum of all these contributions vanishes in the loop, i.e., 
\be
 \sum_{i=1}^4 T^{(22)}_i ( a , a_1 , a_2)  \Big( F^{(22)}_{i,\text{IR }}(\kvec , \qvec) + F^{(22)}_{i,\text{IR }}(\kvec , - \qvec)  \Big)  \\
+ \sum_{i = 1}^6 T^{(13)}_i (a, a_1 , a_2)  F^{(13)}_{i, \text{IR }} ( \kvec , \qvec)    =0 \;,
\label{sumIR}
\ee
as expected. We can use this result to define the IR-safe kernels by subtracting out the divergent contribution from each kernel, i.e.,
 \begin{align}
  F^{(13)}_{i,\text{IR-safe}} (\kvec , \qvec)  & \equiv F^{(13)}_i(\kvec , \qvec) -  F^{(13)}_{i , \text{IR}}(\kvec , \qvec) \, \Theta_{\rm H} ( k - q)  \ , \\
  F^{(22)}_{i,\text{IR-safe}} (\kvec , \qvec) & \equiv F^{(22)}_i(\kvec , \qvec) \, \Theta_{\rm H} ( | \kvec - \qvec| - q )  -  F^{(22)}_{i , \text{IR}}(\kvec , \qvec) \, \Theta_{\rm H} ( k - q) \ . 
 \end{align} 
By virtue of eq.~\eqref{sumIR}, computing the one-loop power spectrum using these redefined kernels does not change the final result, but now that the spurious IR pieces have been removed, the integral can be done with much less precision.

Let us now consider the UV divergences, obtained in the limit $k/q \to 0$.  
In this limit the kernels have the form
\begin{align}
F^{(22)}_i ( \kvec , \qvec )  & =  \mathcal{O} \left( {k^4}/{q^4}  \right) P_{q}^{\rm in} P_{q}^{\rm in}  \qquad i=1,2,3,4\;, \\
F^{(13)}_i ( \kvec , \qvec )  & = \left(  - 2 {\mu^2}     + \mathcal{O} \left( {k^2}/{q^2} \right) \right) P_{k}^{\rm in} P_{q}^{\rm in} \qquad i =1,2 \;, \\
F^{(13)}_i ( \kvec , \qvec )  & = \left(  2 {\mu^2}    + \mathcal{O} \left( {k^2}/{q^2} \right) \right) P_{k}^{\rm in} P_{q}^{\rm in} \qquad i =3,4 \;, \\
F^{(13)}_i ( \kvec , \qvec )  & =  \mathcal{O} \left( {k^2}/{q^2} \right)  P_{k}^{\rm in} P_{q}^{\rm in} \qquad i =5,6 \;.
\end{align}
Conservation of mass and momentum implies that effects from the UV can only start at order $  k^2  P_{k}^{\rm in} $ or $k^4$. These are simply the terms which can be adjusted by counterterms in the EFTofLSS, the former contribution coming from $\partial^2 \delta$, and the latter coming from a stochastic piece \cite{Carrasco:2012cv}. Therefore, the terms proportional to $k^0/q^0$ in $F^{(13)}_i $ must be absent in the final result.\footnote{As noted in \cite{Carrasco:2013sva}, one can also subtract out the terms which are degenerate with the counterterms.  Because these parts of the loop integral will be adjusted by counterterms anyway, one does not have to waste computational time computing them in the loop integrals.} 
Indeed, once can check that 
\be
\sum_{i = 1}^6 T^{(13)}_i (a, a_1 , a_2)  F^{(13)}_{i, \text{UV }} ( \kvec , \qvec)    =0 \;,
\label{sumUV}
\ee
so that the UV divergences cancel in the one-loop power spectrum. Combining with the results obtained above, we can then define the IR\&UV-safe kernels as
\be
\begin{split}
  F^{(13)}_{i,\text{IR\&UV-safe}} (\kvec , \qvec)  & \equiv F^{(13)}_i(\kvec , \qvec) -  F^{(13)}_{i , \text{IR}}(\kvec , \qvec) \, \Theta_{\rm H} ( k - q)  - F^{(13)}_{i , \text{UV}}(\kvec , \qvec) \, \Theta_{\rm H} (  q - k ) \ , \\
  F^{(22)}_{i,\text{IR\&UV-safe}} (\kvec , \qvec) & \equiv F^{(22)}_i(\kvec , \qvec) \, \Theta_{\rm H} ( | \kvec - \qvec| - q )  -  F^{(22)}_{i , \text{IR}}(\kvec , \qvec) \, \Theta_{\rm H} ( k - q) \ , \label{IRUVsafe}
 \end{split}
 \ee 
with 
\be
\begin{split}
 F^{(13)}_{1 , \text{UV}} ( \kvec , \qvec) &=  F^{(13)}_{2, \text{UV}} ( \kvec , \qvec) =  - F^{(13)}_{3, \text{UV}} ( \kvec , \qvec) =  - F^{(13)}_{4, \text{UV}} ( \kvec , \qvec)  =  - 2 \mu^2   P^{\rm in}_{\kvec} P^{\rm in}_{\qvec} \ , \\
 F^{(13)}_{5 , \text{UV}} ( \kvec , \qvec) &=  F^{(13)}_{6, \text{UV}} ( \kvec , \qvec) = 0  \ . \label{UVcont} 
 \end{split}
 \ee
 Using these kernels instead of the original ones for the computation of the one-loop power spectrum removes both spurious IR and UV divergences.

Let us now discuss  the non-standard terms. We must thus consider the kernels $F^{(22)}_i ( \kvec , \qvec )$ for $i=5,6,7$ and $F^{(13)}_i ( \kvec , \qvec)$ for $i=7,\dots,11$. Starting from the IR divergences,
in the limit $q/k \to 0$ we have
\begin{align}
\begin{split}
F^{(22)}_5 ( \kvec , \qvec ) & = \left(   \mu \left( 1 - \mu^2 \right) \frac{k}{q} + \mathcal{O} \left( {k^0}/{q^0} \right) \right) P_{k}^{\rm in} P_{q}^{\rm in}  \;, \\
F^{(22)}_6 ( \kvec , \qvec ) & =  \left(   \mu \left( 1 - \mu^2 \right) \frac{k}{q} + \mathcal{O} \left( {k^0}/{q^0} \right) \right) P_{k}^{\rm in} P_{q}^{\rm in} \;, \\
F^{(22)}_7 ( \kvec , \qvec ) & = \mathcal{O} \left( {k^0}/{q^0} \right)  P_{k}^{\rm in} P_{q}^{\rm in}  \ ,
\end{split}
\end{align}
and 
\be
F^{(13)}_i ( \kvec , \qvec)  =  \mathcal{O} \left( {k^0}/{q^0} \right)   P_{k}^{\rm in} P_{q}^{\rm in}  \qquad i=7,\ldots,11 \;. 
\ee
There are no divergences in $k^2/q^2$ neither for $q/k \to 0$ nor for $\qvec \rightarrow \kvec$.
To treat the divergences linear in $k/q$ in $F^{(22)}_i$, it is sufficient to adopt the same procedure for the standard terms outlined by eq.~\eqref{rewritingkq}. The kernels $F^{(13)}_i $ have no IR divergences in the non-standard case, so we do not have to make any modifications of the integrand.

We can turn to the UV limit. For $ k / q \rightarrow 0$ we have
\begin{align}
F^{(22)}_i ( \kvec , \qvec) & = \mathcal{O} \left(  {k^4}/{q^4} \right) P_{q}^{\rm in} P_{q}^{\rm in} \qquad i=5,6,7 \;,  \\
F^{(13)}_i ( \kvec , \qvec) & = \mathcal{O} \left(  {k^2}/{q^2} \right)  P_{k}^{\rm in} P_{q}^{\rm in} \qquad i= 7,\ldots,11 \ , 
\end{align}
which are simply contributions  degenerate with the counterterms. Therefore, we do not have to make any subtractions to make the integrand UV-safe.  
As mentioned before, we could choose to save computational time by subtracting these terms out of the loops.  However, because the gain is not very significant in the one-loop calculation, we choose for simplicity not to subtract out the above pieces.

To conclude, to compute the IR\&UV-safe one-loop power spectrum we can simply use the results of \cite{Lewandowski:2017kes} valid for the standard case without modifications of gravity, i.e.~the kernels defined in eq.~\eqref{IRUVsafe}, where the non-vanishing IR  and UV contributions are given respectively by eqs.~\eqref{IRcont} and \eqref{UVcont}.


\newpage

 \bibliographystyle{utphys}
\bibliography{EFT_DE_biblio3}

\end{document}